\documentclass[twocolumn]{emulateapj}
\usepackage{apjfonts}

\usepackage{graphicx}
\usepackage{mathrsfs}
\usepackage{natbib}
\bibpunct[,]{(}{)}{;}{a}{}{,}

\def\totd{{\mathrm{d}}}
\def\rs0{{r_{\rm s0}}}
\def\tff0{{t_{\rm ff0}}}
\def\vff2{{v_{\rm ff0}^2}}

\shorttitle{3D SASI}
\shortauthors{Fern\'andez}

\slugcomment{Submitted 2010 March 6; Accepted 2010 October 9}

\begin{document}

\title{The Spiral Modes of the Standing Accretion Shock Instability}
\author{Rodrigo Fern\'andez\altaffilmark{1}}
\affil{Institute for Advanced Study. Einstein Drive, Princeton, NJ 08540, USA.}
\altaffiltext{1}{Einstein Fellow}

\begin{abstract}
A stalled spherical accretion shock, such as that arising in core-collapse supernovae, is unstable 
to non-spherical perturbations. In three dimensions, this Standing Accretion Shock Instability (SASI)
can develop spiral modes that spin-up the protoneutron star. 
Here we study these non-axisymmetric modes by combining linear stability analysis and three-dimensional, 
time-dependent hydrodynamic simulations with Zeus-MP, 
focusing on characterizing their spatial structure and angular momentum content.
We do not impose any rotation on the background accretion flow, and use simplified microphysics with no neutrino 
heating or nuclear dissociation.  
Spiral modes are examined in isolation by choosing flow parameters such that only the fundamental mode is
unstable for a given polar index $\ell$, leading to good agreement with linear theory.
We find that any superposition of sloshing modes with non-zero relative phases survives in the nonlinear regime and leads
to angular momentum redistribution. It follows that the range of perturbations required to obtain spin-up
is broader than that needed to obtain the limiting case of a phase shift of $\pi/2$.
The bulk of the angular momentum redistribution occurs during a phase of exponential growth, and arises
from internal torques that are second order in the perturbation amplitude. This redistribution gives rise
to at least two counter rotating regions, with the maximum angular momentum of a given sign approaching a 
significant fraction of the mass accretion rate times the shock radius squared 
$(\dot{M}\,r_{\rm shock}^2\sim 10^{47}$~g~cm$^2$~s$^{-1}$, spin period $\sim 60$~ms). Nonlinear mode coupling at 
saturation causes the angular momentum to fluctuate in all directions with much smaller amplitudes.
\end{abstract}


\maketitle

\section{Introduction}

The explosion mechanism of massive stars is not well understood at present.
Observational and theoretical evidence gathered thus far suggests that this mechanism is 
intrinsically asymmetric for all stars that form iron cores ($\gtrsim 12M_\sun$; see, e.g., 
\citealt{wang08}, \citealt{burrows07d} and \citealt{janka07} for recent reviews). Stars that end up with  O-Ne-Mg 
cores are currently found to explode in spherical symmetry via the \emph{neutrino mechanism}
(\citealt{kitaura06}, \citealt{burrows07c}, \citealt{janka08}).

One piece of phenomenology that a successful theory of core-collapse supernovae has to 
explain is the distribution of pulsar spins at birth. Population synthesis studies of radio pulsars
generally assume normal or log-normal distributions with mean values ranging 
from a few ms \citep{arzoumanian02} to a few $\sim 100$~ms \citep{FG-K06} in order
to reproduce observations. 
This large range is due to the different
assumptions made about input physics, such as the shape of the 
radio beam (e.g., \citealt{FG-K06}). Independent constraints that combine \emph{Chandra}, \emph{XMM}, 
and \emph{Swift} observations of
historic supernovae with an empirical correlation between X-ray luminosity and spin down power tend
to rule out a large pulsar population with initial periods $\leq 40$~ms \citep{perna08}.
On the other hand, current stellar evolution calculations that account for magnetic 
torques predict, using conservation of angular momentum and an assumption for the mass cut, 
neutron star birth periods $\lesssim 10$~ms \citep{heger05}. 
However, these models suffer from large uncertainties in the
input physics, which could also lead in principle to very slowly rotating pulsars \citep{spruit98}. 
Axisymmetric core-collapse calculations indicate that there is an almost linear mapping between
the rotation rate of the iron core and that of the resulting protoneutron star, with no robust
braking mechanism in sight to bring the implied fast neutron star spins to values more
in agreement with observations \citep{ott06}.

If most presupernova cores turn out to rotate slowly, then there is an alternative mechanism
to generate periods $\gtrsim 50$~ms, which arises from instabilities
in the supernova shock itself \citep{blondin07a}. Axisymmetric core-collapse supernova
simulations that include neutrino transport and microphysics to several degrees
of approximation find that the stalled postbounce shock undergoes large scale oscillatory motions 
that break the spherical symmetry of the system \citep{burrows95,janka96,mezzacappa98,scheck06,ohnishi06,
buras06a,buras06b,burrows06a,burrows07,
scheck08,murphy08,ott08,marek09,suwa09}. When neutrino driven convection 
is suppressed, the instability of the shock persists in the form
of an overstable sloshing cycle, with the most unstable modes having Legendre 
indices $\ell=1$ and $2$ \citep{BM03,BM06}. In the limit of short wavelength, this so-called
Standing, Spherical, or Stationary Accretion Shock Instability (SASI) is driven by an interplay 
between advected and acoustic perturbations \citep{F07}. There is no conclusive proof yet on the
mechanism behind long wavelength modes, although considerations of the timescales \citep{FT09a} 
and the case of planar geometry \citep{foglizzo09,sato09} 
point to the advective-acoustic cycle as a relevant component.
The SASI is weakened when rotation becomes dynamically important (e.g., \citealt{ott08}).

Three-dimensional simulations of the SASI have found that a spiral type of mode
can develop, causing the flow to divide itself into at least two counter-rotating streams,
leading to angular momentum redistribution
\citep{blondin07a}. 
\citet{blondin07a} find that this process can spin-up a canonical neutron
star to a period of $\sim 50$~ms after $\sim 0.1M_\sun$ of material is accreted. 
Aside from the fact that this spin period is comparable
to that obtained by population synthesis calculations, the most interesting
results of \citet{blondin07a} are that (i) no progenitor rotation is required for
this mechanism to operate and, (ii) for weak progenitor rotation, this instability
dominates the spin-up of the protoneutron star, imparting it along a different axis.

One of the key questions that remains to be answered is how easy it is to excite
these spiral modes. In contrast to \citet{blondin07a} and \citet{blondin07b}, the 
findings of \citet{iwakami08}, \citet{iwakami09a}, and 
\citet{iwakami09b} suggest that with no 
rotation in the upstream flow, these modes are difficult to excite.
\citet{yamasaki08} find that, indeed, for a cylindrical accretion flow with
rotation, prograde spiral modes have a higher growth rate than retrograde ones.
They attribute this to a Doppler shift of the mode frequency induced by rotation. 
Part of the difficulty in comparing numerical results on this instability
is that they have been obtained using different microphysics, numerical methods, and coordinate
systems, and that the three-dimensional structure of these linear eigenmodes remains as of yet unexplored.
The work of \citet{blondin07b} examined spiral SASI modes with time-dependent
hydrodynamical simulations restricted to a polar wedge around the equatorial plane.
They found that sloshing and spiral modes are closely related, because the pressure
perturbation rotates with the same frequency as a sloshing mode of the same Legendre
index, and because by combining two counter-rotating spirals, the flow field of a sloshing
mode is recovered. The work of \citet{iwakami09b} also identified sloshing modes as the superposition
of two counter-rotating spirals.

The aim of this paper is to better understand spiral modes in the linear and nonlinear phase, particularly
their three-dimensional spatial structure and angular momentum content.

The structure of axisymmetric sloshing eigenmodes has been obtained previously 
through linear stability analysis \citep{F07}, and verified to high
precision with time-dependent axisymmetric hydrodynamic simulations \citep{FT09a}.
Starting from the findings of \citet{blondin07b}, we show that spiral modes are
most easily understood as sloshing modes out of phase. Their spatial structure is 
built using solutions to the differential system of \citet{F07}, and their evolution is compared
with results of time-dependent hydrodynamic simulations using Zeus-MP \citep{hayes06}.
In our calculations we do not add rotation to the flow anywhere, and employ
an ideal gas equation of state, parametric cooling, and point-mass gravity,
with no heating or nuclear dissociation below the shock. 

One of our main findings is that it is relatively simple to create spiral modes out of
sloshing modes by changing their numbers, relative phases, and amplitudes, covering
a broader parameter space than the limiting cases where the amplitudes are equal and the
relative phase is $\pi/2$. These modes survive to large
amplitudes, resulting in a significant angular momentum redistribution below the shock. 
We set aside for now
the question of whether this coherent superposition of sloshing modes is
able to develop in a more realistic core-collapse context, where three-dimensional,
nonlinear turbulent convection may likely act as a forcing agent \citep{FT09b}.
We revisit this issue in the discussion section.

We also find that the bulk of the angular momentum redistribution
generated by a spiral mode occurs during the phase of exponential growth.
This spin-up of the flow arises from internal torques that are second order
in the perturbation amplitude, and results in angular momenta of a 
characteristic magnitude at saturation. Non-linear mode coupling causes the 
angular momentum flux to become stochastic, fluctuating
in all directions with an amplitude much smaller than that achieved during 
exponential growth.

The structure of this paper is the following. In Section 2 we describe the physical model used,
and summarize the most important elements that enter the linear
stability calculation and time-dependent simulations, with details deferred to the Appendices. 
Section 3 describes how spiral modes can be constructed by superposing sloshing modes out of phase,
and explores some of their features.
Section 4 contains the results of time-dependent simulations of these modes, including both
linear and nonlinear development. We conclude by summarizing our findings and discussing their 
implications for more realistic core-collapse models and neutron star spins.

\section{Methods}

\subsection{Physical Model}
\label{sec:phys_model}

The physical system consists of a steady-state, standing spherical accretion shock
at a radial distance $\rs0$ from the origin. Below the shock, the material
settles subsonically onto a star of radius $r_*$ and mass $M$ centered at the origin of the
coordinate system. This settling is mediated by a cooling rate per unit volume
that mimics neutrino emission due to electron capture,
$\mathscr{L} \propto p^{3/2}\,\rho$, with $p$ the pressure and $\rho$ the density 
(e.g., \citealt{BM06}, \citealt{FT09a}).
In all the cases studied here, cooling is relevant only in a narrow region -- of
order a pressure scale height -- outside the accretor. Upstream of the shock, the fluid is supersonic, 
with incident Mach number $\mathcal{M}_1 = 5$ at $r=\rs0$, and has a vanishing energy flux.
Upstream and downstream solutions are connected by the Rankine-Hugoniot jump
conditions.
The equation of state is that of an ideal gas of adiabatic index $\gamma=4/3$,
and the self gravity of the flow is neglected. The equations describing this
steady flow are presented in \S\ref{s:background}.
The model was first developed by \citet{HC92} to study
accretion onto compact objects, and subsequently used by 
\citet{BM03}, \citet{BM06}, \citet{F07}, \citet{blondin07a}, \citet{blondin07b}, and
\citet{FT09a} as a minimal approximation to the postbounce stalled shock.

In contrast to \citet{FT09a}, we do not include here the effects of nuclear
dissociation at the shock. This is a significant energy sink in the system,
changing the Mach number and density below the shock
for fixed upstream conditions, and hence affecting the linear modes of the SASI.
Nevertheless, as our goal here is to characterize the basic behavior of the
instability in three dimensions, we aim at keeping the calculations as simple
as possible.

Neutrino heating is also neglected, in order to suppress
convection in the flow, as originally done by \citet{BM03}. Since most of the flow below
the shock is adiabatic, the entropy profile is flat and hence marginally
unstable to overturns triggered by the SASI itself.

We adopt a system of units normalized to the shock radius $\rs0$, the free-fall
velocity at this radius $v_{\rm ff0} = \sqrt{2GM/\rs0}$, and the upstream density $\rho_1$ \citep{FT09a}.
Numerical values relevant for the stalled shock phase of core collapse supernovae are $r_\mathrm{s0} \sim 150$~km,
$v_\mathrm{ff}(r_{\rm s0}) \sim 4.8\times 10^9 M_\mathrm{1.3}^{1/2}\,(r_{\rm s0}/150~{\rm km})^{-1/2}$~km s$^{-1}$,
$t_\mathrm{ff} \equiv r_\mathrm{s0}/v_\mathrm{ff} \sim 3.1 \,M_\mathrm{1.3}^{-1/2}\,
(r_{\rm s0}/150~{\rm km})^{3/2}$~ms, and
$\rho_1 \sim 4.4\times 10^7 \dot{M}_\mathrm{0.3}\,M_\mathrm{1.3}^{-1/2}\,(r_{s0}/150~{\rm km})^{-3/2}$~g
cm$^{-3}$ (assuming a strong shock), where the mass accretion rate has been normalized to
$\dot M = 0.3\,\dot{M}_\mathrm{0.3}\,M_\odot$ s$^{-1}$. We will express quantities involving 
angular momentum in terms of $\dot{M}\rs0 \simeq 4\pi\times 10^{46}(r_{\rm s0}/150~{\rm km})^2 \dot{M}_\mathrm{0.3}$~g~cm$^2$~s$^{-1}$.

\subsection{Linear Stability Analysis}
\label{sec:linear_stability}

To obtain linear SASI eigenmodes and eigenfrequencies, we solve the differential 
system of \citet{F07}, the results of which agree with time-dependent
axisymmetric hydrodynamic simulations to high precision \citep{FT09a}.
The relevant equations are described in detail in Appendix~\ref{sec:v_trans}; here we
summarize the main elements of the calculation.
The eigenvalue problem is formulated in terms of perturbation variables
that are conserved when advected. These are the perturbed mass flux $h = \delta\rho/\rho +\delta v_r/v_r$, 
Bernoulli flux $f = v_r\delta v_r + \delta c^2/(\gamma-1)$,
entropy $\delta S = (\gamma-1)^{-1}\left[\delta p/p - \gamma\delta \rho/\rho\right]$, 
and an entropy-vortex combination $\delta K = r^2\mathbf{v}\cdot(\nabla\times\delta\mathbf{w}) +
\ell(\ell+1)(p/\rho)\delta S$ (see also eq.~[\ref{eq:K_redef}]),
with $c^2 = \gamma p/\rho$, $\mathbf{v}$ the velocity, and $\mathbf{w}$ the vorticity.
Perturbations are decomposed in Fourier modes in time and spherical harmonics
in the angular direction, having the general form 
\begin{equation}
\label{eq:perturbation_form}
\delta q = \delta \tilde{q}(r)\,e^{-i\omega t}\,Y_\ell^m(\theta,\phi),
\end{equation}
where $\delta \tilde{q}(r)$ is the complex amplitude, $Y_\ell^m$ is a spherical harmonic, and the complex 
frequency satisfies $\omega = \omega_{\rm{osc}} + i\omega_{\rm{grow}}$. 
This choice results in all thermodynamic
variables along with $\delta v_r$ being proportional to $Y_\ell^m$ (corresponding to \emph{spheroidal} modes,
e.g. \citealt{andersson01}).
The transverse components of the velocity
involve angular derivatives of the $Y_\ell^m$, and have the same radial amplitude (Appendix~\ref{sec:v_trans}),
\begin{eqnarray}
\delta v_\theta & = & \delta \tilde{v}_\Omega(r)\,e^{-i\omega t}\,\frac{\partial}{\partial \theta}Y_\ell^m(\theta,\phi)\\
\delta v_\phi   & = & \delta \tilde{v}_\Omega(r)\,e^{-i\omega t}\, 
                      \frac{1}{\sin\theta}\frac{\partial}{\partial \phi}Y_\ell^m(\theta,\phi).
\end{eqnarray}

The system of ordinary differential equations that determines
the complex amplitudes is independent of the azimuthal number $m$ of the mode. 
Boundary conditions at the shock
are expressed in terms proportional to the shock displacement $\Delta \xi$, or the shock velocity 
$\Delta v = -i\omega \Delta \xi$. By imposing $\delta \tilde{v}_r(r_*) = 0$ at the inner boundary, a
complex eigenvalue $\omega$ is obtained for a given set of flow parameters. For each
$\ell$, a discrete set of overtones is obtained, which are related to the number of nodes
in the radial direction \citep{F07,FT09a}. The resulting eigenmodes have both real
and imaginary eigenfrequencies.

For fixed $\gamma$, $\mathcal{M}_1$, and functional form of the cooling function, the
eigenfrequencies of the SASI depend only on the relative size of the shock
and accreting star, quantified by the ratio $r_*/\rs0$.
Time-dependent studies of individual SASI modes are possible by choosing 
model parameters such that only the fundamental mode is unstable, for a given $\ell$.
Based on the fact that the relevant modes in more realistic core-collapse simulations
are $\ell =1$, and $2$, we adopt two configurations for single-mode studies: $r_*/\rs0 = 0.5$ for $\ell=1$, 
and $r_*/\rs0 = 0.6$ for $\ell=2$ (see \citealt{F07} and \citealt{FT09a} for more extended
parameter studies of the SASI eigenfrequencies). In addition, we explore a configuration
with a larger postshock cavity, $r_*/\rs0 = 0.2$, to more closely resemble actual stalled 
supernova shocks. In this case unstable overtones of $\ell=1$ and $\ell=2$ are present.

To facilitate visualization, and unless otherwise noted, we employ the \emph{real} representation of spherical harmonics.
For $\ell=1$, one has
\begin{eqnarray}
\label{eq:Y_z}
Y_1^0    & = & \sqrt{\frac{3}{4\pi}}\cos\theta\\
\label{eq:Y_x}
Y_1^1    & = & \sqrt{\frac{3}{4\pi}}\sin\theta\,\cos\phi\\
\label{eq:Y_y}
Y_1^{-1} & = & \sqrt{\frac{3}{4\pi}}\sin\theta\,\sin\phi.
\end{eqnarray}
This basis is entirely equivalent to the one involving complex exponentials in azimuth (e.g., \citealt{arfken05}),
but has a straightforward interpretation: equations~(\ref{eq:Y_z}), (\ref{eq:Y_x}), and (\ref{eq:Y_y})
are dipoles in the z-, x-, and y-directions, respectively. 
For $\ell=2$, we have
\begin{eqnarray}
Y_2^0    & = & \sqrt{\frac{5}{4\pi}}\frac{1}{2}\left(3\cos^2\theta-1\right)\\
Y_2^1    & = & \sqrt{\frac{15}{4\pi}}\sin\theta\cos\theta\,\cos\phi\\
Y_2^{-1} & = & \sqrt{\frac{15}{4\pi}}\sin\theta\cos\theta\,\sin\phi\\
Y_2^2    & = & \sqrt{\frac{15}{16\pi}}\sin^2\theta\, \cos 2\phi\\
\label{eq:Y_2-2}
Y_2^{-2} & = & \sqrt{\frac{15}{16\pi}}\sin^2\theta\, \sin 2\phi,
\end{eqnarray}
corresponding to the usual z-symmetric quadrupole ($Y_2^0$), and a series of 4-striped "beach balls" with
alternating polarity and symmetry axis along $\hat x$ ($Y_2^1$), $\hat y$ ($Y_2^{-1}$), and $\hat z$ ($Y_2^{\pm 2}$). 

Whenever necessary, we will denote the traditional \emph{complex spherical harmonics} as $\Upsilon_\ell^m$.

\subsection{Time Dependent Numerical Simulations}
\label{sec:time_dep_setup}

We perform time-dependent hydrodynamic calculations to verify the
evolution of linear SASI eigenmodes in three dimensions, covering the
linear and nonlinear phases. 
To this end, we employ the publicly available code Zeus-MP \citep{hayes06},
which solves the Euler equations using a finite difference algorithm that includes
artificial viscosity for the treatment of shocks. The default version of the code
supports an ideal gas equation of state and point mass gravity. We
have extended it to account for optically thin cooling and a tensor artificial
viscosity, for a better shock treatment in curvilinear
coordinates (e.g., \citealt{stone92}, \citealt{hayes06}, \citealt{iwakami08}).
Issues associated with implementing the latter are discussed in Appendix~\ref{sec:zeus}.

The initial conditions are set by the steady-state accretion flow used in linear
stability calculations. To prevent runaway cooling at the base of the flow,
we impose a gaussian cutoff in the cooling with entropy \citep{FT09a}. The normalization
of the cooling function is adjusted so that the Mach number at the surface of the
accreting star is $\mathcal{M}\sim 10^{-2}$. 

Spherical polar coordinates ($r,\theta,\phi$) are used , covering the whole sphere minus
a cone of half-opening angle $5$ degrees around the polar axis. This prescription
does not significantly alter the flow dynamics, and ameliorates the severe Courant-Friedrichs-Lewy restriction
around the polar axis (H-Th. Janka, private communication). We present convergence studies
in Appendix~\ref{sec:zeus} showing that only $\sim 10\%$ differences relative to using
the full sphere are introduced with this prescription.
Cells are uniformly spaced in the polar and azimuthal
directions, while ratioed in radius. The choice of grid spacing is key to
a achieve a stable background solution and to adequately capture linear
growth rates, while minimizing the computational cost. The radial size of the cells at $r=r_*$, 
$\Delta r_{\rm min}$, determines how stable the unperturbed shock remains, because near hydrostatic equilibrium 
needs to be maintained. 
At the shock, the angular size determines how well growth rates are captured. 
Based on numerical experiments, we have found that 
$\Delta r_{\rm min} < 10^{-3}\rs0$ is required to obtain a stable shock within $500$ dynamical times,
and that 36 cells in the polar direction is the minimum needed to obtain a clean
shock oscillation ($\omega_{\rm osc}$ within $\lesssim 10\%$ and $\omega_{\rm grow}$ within $\sim 20\%$ from the
linear stability value; convergence tests are presented in Appendix~\ref{sec:zeus}). 
We choose the ratio of radial spacing so that $\Delta r = \rs0 \Delta \theta$ 
at the shock radius $\rs0$, and $\Delta \phi = \Delta \theta$, resulting in a total of $56$ radial ($\ell=1$),
$48$ polar, and $96$ azimuthal cells. For $\ell=2$, we use $28$ ratioed cells in radius from $r_*$ to $\rs0$, and
then use a uniform radial spacing until the outer boundary. This choice of grid spacing is made because the tensor 
artificial viscosity behaves best when applied over the longest cell dimension (e.g., \citealt{stone92}), and 
hence choosing the three cell dimensions as equal as possible results in nearly isotropic dissipation
except near the polar axis\footnote{The convergence of meridional grid lines inevitably results in a much 
smaller cell size in the azimuthal direction next to the shock.}.
In addition, we perform one simulation at double the resolution in all dimensions (e.g., 128x96x192) 
to test the robustness of our results. This resolution approaches values comparable to existing two-dimensional
simulations, and can better capture the parasitic instabilities that mediate the saturation
of the instability \citep{guilet09a}. In Table~\ref{t:models} we summarize the modes evolved. With the exception of the
high-resolution model (R5\_L1P2\_HR), all simulations were carried out until at least $200\tff0$, well into the nonlinear phase.
\begin{deluxetable}{lcll}
\tablecaption{Models Evolved.\label{t:models}}
\tablewidth{0pt}
\tablehead{
\colhead{Model\tablenotemark{a}} &
\colhead{$r_*/\rs0$} &
\colhead{Perturbation\tablenotemark{b}} &
\colhead{Resolution\tablenotemark{c}}
}
\startdata
r5\_L1\_LR(c) & 0.5 & Shell, $\ell=1$       &  $\phantom{1}56\times 48$              \\
r5\_L1\_HR(c) & 0.5 & Shell, $\ell=1$       &  $          112\times 96$              \\
\noalign{\smallskip}
R2\_L11f(c)    & 0.2 & Shells, $\Phi=\pi/2$  &  $\phantom{1}56\times 48\times 96$    \\
R2\_L11h(c)    & 0.2 & Shells, $\Phi=\pi/2$  &  $\phantom{1}56\times 48\times 96$    \\
\noalign{\smallskip}
R5\_L11x       & 0.5 & Shell,  $Y_1^1$ only  &  $\phantom{1}56\times 48\times 96$    \\
R5\_L11P2(c)   & 0.5 & Shells, $\Phi=\pi/2$  &  $\phantom{1}56\times 48\times 96$    \\
R5\_L11P4      & 0.5 & Shells, $\Phi=\pi/4$  &  $\phantom{1}56\times 48\times 96$    \\
R5\_L11P8      & 0.5 & Shells, $\Phi=\pi/8$  &  $\phantom{1}56\times 48\times 96$    \\
R5\_L11\_HR    & 0.5 & Shells, $\Phi=\pi/2$  &  $          112\times 96\times 192$   \\
R5\_RAN        & 0.5 & Random Pressure 1\%   &  $\phantom{1}56\times 48\times 96$    \\ 
\noalign{\smallskip}
R6\_L21P2(c)  & 0.6 & Shells, $\Phi=\pi/2$  &  $(28+44)\times 48\times 96$    \\
R6\_L21P4     & 0.6 & Shells, $\Phi=\pi/4$  &  $(28+44)\times 48\times 96$    \\
R6\_L21P8     & 0.6 & Shells, $\Phi=\pi/8$  &  $(28+44)\times 48\times 96$    \\
R6\_L22P2(c)  & 0.6 & Shells, $\Phi=\pi/2$  &  $(28+44)\times 48\times 96$    \\
R6\_L22P4     & 0.6 & Shells, $\Phi=\pi/4$  &  $(28+44)\times 48\times 96$    \\
R6\_L22P8     & 0.6 & Shells, $\Phi=\pi/8$  &  $(28+44)\times 48\times 96$    
\enddata
\tablenotetext{a}{In three-dimensional runs, we denote by L$\ell$|m| the case where both $Y_\ell^{\pm m}$ modes are excited
with the quoted phase difference $\Phi$. The symbol (c) indicates that, in addition to the 5 degree cutout around the polar axis,
a run with the full range of polar angles was performed in the linear phase only. In this case, the letter c is appended to the
model name to denote the version \emph{with} cutout. The letters $f$ and $h$ 
mean that either the fundamental, or first overtone was excited, respectively.}
\tablenotetext{b}{See eq.~(\ref{eq:trigger_eqn}) for the definition of $\Phi$.}
\tablenotetext{c}{Number of cells in the $r$, $\theta$, and $\phi$ directions, respectively. For the $r_*/\rs0 = 0.6$ runs,
we use a ratioed grid for $r\leq \rs0$, and thereafter keep the radial cell spacing constant.
}
\end{deluxetable}

The boundary conditions are reflecting at the inner radial boundary ($r=r_*$), periodic in the $\phi$ direction,
reflecting at the polar boundaries ($\theta = \{0,\pi\}$ minus the cutout), and set equal to the upstream flow at the outer 
radial boundary ($r = 4\rs0$). 
The use of tensor artificial viscosity prevents the appearance of a numerical instability at the poles
of the shock, as described by \citet{iwakami08b}. This \emph{carbuncle} instability \citep{quirk94} arises 
within a few dynamical times if the standard \citet{vonneumann50} prescription is used.  
However, even when including the tensor viscosity, we have found that a different numerical 
instability develops around the polar axis when reflecting boundary conditions and the full sphere are used. 
This is due to the fact that a wake is generated by the reflecting axis when the sloshing modes reach a large
amplitude, resulting in a different radial velocity profile at opposite sides of the polar axis. 
The azimuthal velocity then develops 
a sawtooth instability around the axis. The cutout improves (but does not completely suppress)
the development of this instability, to the point where it does not significantly alter the flow dynamics.

\begin{figure*}
\includegraphics*[width=\textwidth]{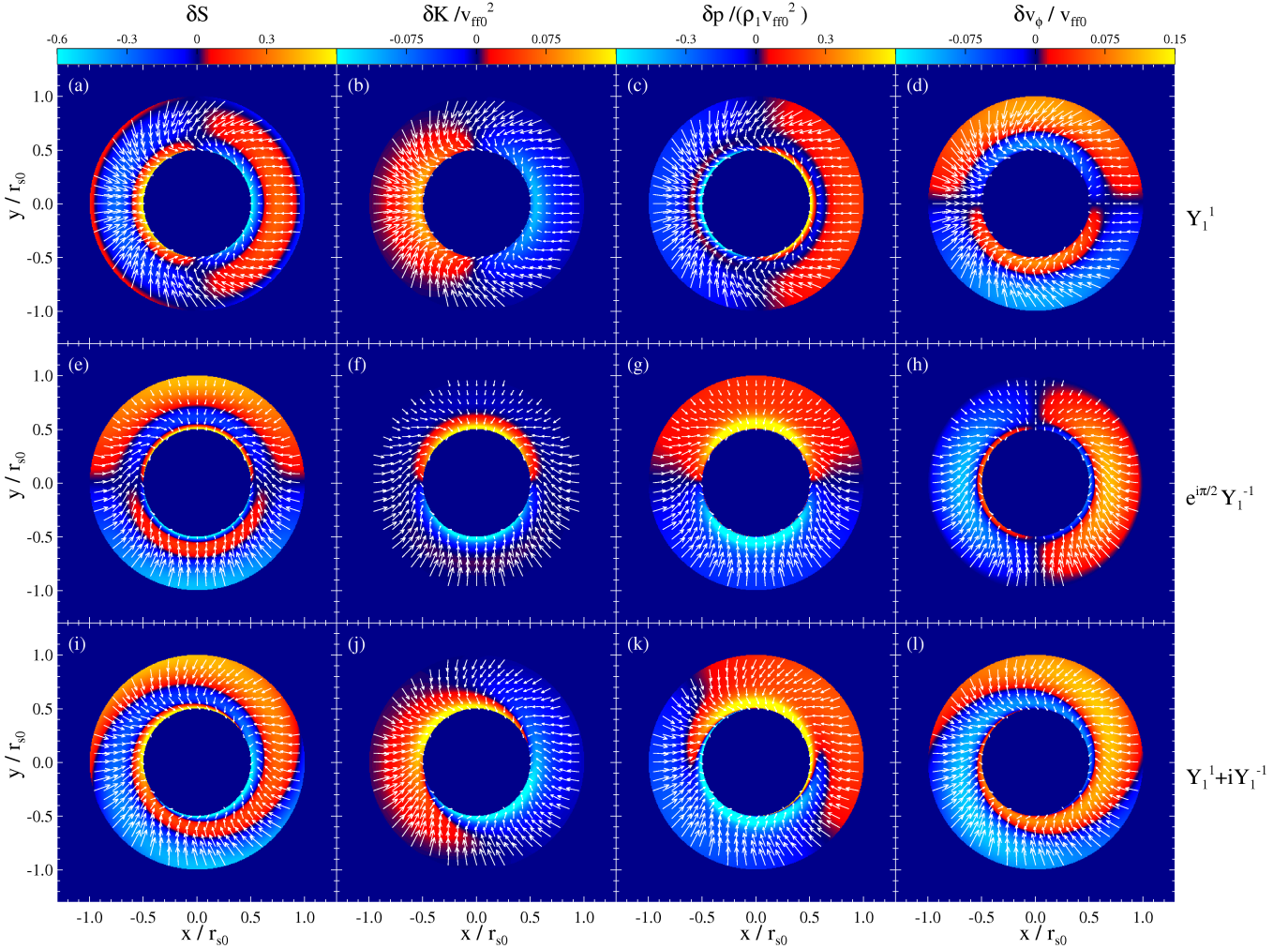}
\caption{Real part of perturbed quantities from linear stability analysis, showing how transverse sloshing modes
out of phase combine to create a $\ell=1$, $m=1$ spiral mode at the equatorial 
plane ($z=0$). The top, middle,
and bottom rows show the $Y_1^{1}$ mode at $t=0$, the $Y_1^{-1}$ mode displaced $\pi/2$ in phase, and their 
sum,
respectively (real spherical harmonics are given by eqns.~[\ref{eq:Y_z}]-[\ref{eq:Y_y}]). 
From left to right, columns show the 
entropy, entropy-vortex, pressure, and azimuthal velocity
perturbation for a shock displacement amplitude $\Delta \tilde{\xi} = 0.25\rs0$ (\S\ref{sec:linear_stability} 
and Appendix~\ref{sec:v_trans}). White arrows show the
perturbed velocity profile. The size of the postshock cavity is $r_*/\rs0 = 0.5$, and
only the fundamental mode is unstable. Note that only the region below the unperturbed shock
position is shown. The electronic version of the article contains an animated version of this
figure (real frequency only).}
\label{f:figure_spiral}
\end{figure*}

\begin{figure}
\includegraphics*[width=\columnwidth]{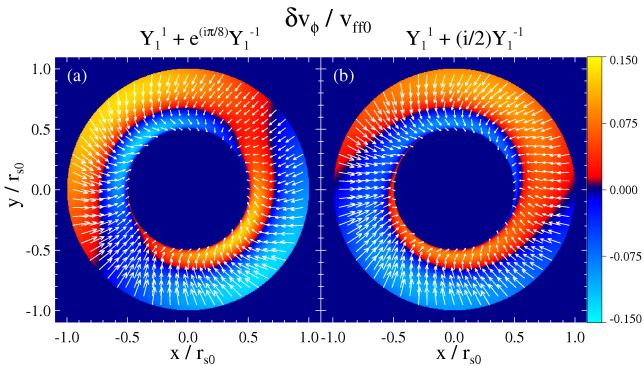}
\caption{Effect of changing the relative phase (left) and amplitude (right) relative to 
the $\ell=1$, $m=1$ spiral in Figure~\ref{f:figure_spiral}l.
Shown is the perturbed azimuthal velocity from the linear eigenmodes at $z=0$. The left panel 
has a phase difference $\Phi = \pi/8$ with equal amplitudes, while the right panel decreases the amplitude
of $Y_1^{-1}$ by 1/2, with a phase difference of $\pi/2$. The electronic version of the article contains
an animated version of this figure (real frequency only).}
\label{f:figurepars}
\end{figure}

\section{Superposition of Linear Eigenmodes}
\label{sec:eigenmode}

Understanding of spiral SASI modes becomes easier when the \emph{real} representation of 
spherical harmonics is used (eqns.~[\ref{eq:Y_z}]-[\ref{eq:Y_2-2}]). 
For $\ell=1$, they correspond to the familiar sloshing modes, axisymmetric relative to the 
three cartesian coordinate axes. When viewed this way, the fact that the eigenfrequencies
for a given $\ell$ are independent of azimuthal number $m$ is natural, as any of these elementary 
modes is equally likely to arise in a spherically symmetric accretion flow (see, e.g., \citealt{binney08} for
a more rigorous argument). When excited in phase, 
any linear combination of these sloshing modes will result in a new sloshing mode that is symmetric 
around some axis. 

\begin{figure*}
\includegraphics*[width=\textwidth]{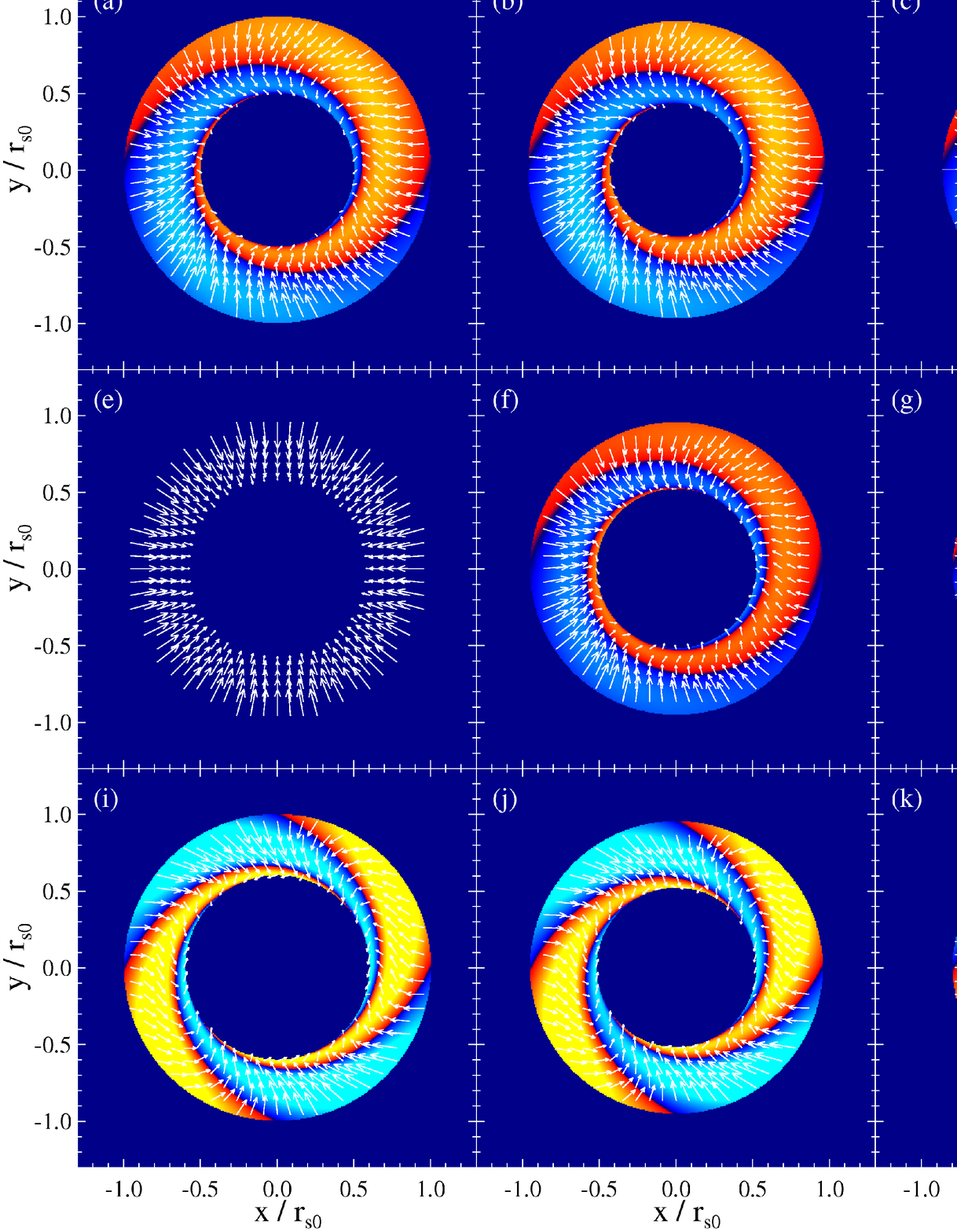}
\caption{Real part of the perturbed azimuthal velocity from linear stability, showing the three-dimensional
structure of spiral modes for different combinations of $\ell$ and $m$. The top, middle,
and bottom rows show $Y_1^{1}+iY_1^{-1}$, $Y_2^1+iY_2^{-1}$, and $Y_2^2+iY_2^{-2}$, all at $t=0$,
respectively. From left to right, columns show different elevations above the equatorial plane.
The fractional stellar size is $r_*/\rs0 = 0.5$ for $\ell=1$ and $r_*/\rs0 = 0.6$ for $\ell=2$, 
so that only the fundamental mode is unstable. The equatorial plane of the $Y_2^{\pm 1}$ mode (panel e)
has only radial flow.}
\label{f:figure_modes} 
\end{figure*}

Spiral modes arise whenever two or more of these ``basis" sloshing modes are excited out of phase.
The familiar $\ell=1$, $m=\pm 1$ spirals corresponds to $Y_1^1 \pm iY_1^{-1}$, or x- and y-symmetric sloshing modes 
that are $\pm \pi/2$ out of phase with each other, and that have the same amplitude, akin to
two transverse and out-of-phase harmonic oscillators describing circular motion.
Aside from the exponential growth in amplitude, this gives rise to a static spiral 
pattern that rotates with the angular frequency of 
the oscillators. This type of spiral motion was suggested by \citet{lin64} to explain the morphology
of spiral arms in galaxies, and is also well-known in stellar pulsation theory (e.g., \citealt{unno89}).

Figure~\ref{f:figure_spiral} shows, side-by-side, the real part of the
entropy, entropy-vortex, pressure, and azimuthal velocity perturbation in the equatorial plane ($z=0$),
with the total velocity field superimposed. Shown are the fundamental modes $Y_1^{-1}$, 
$\exp(i\pi/2)\,Y_1^{-1}$, and the sum of them, for a shock displacement amplitude $\Delta \tilde{\xi} = 0.25\rs0$
(an animated version of the figure is available in the online version of the article). 
In sloshing modes, the entropy eigenmodes consist of annular regions of alternating polarity that 
extend over half a circle in azimuth for $\ell = 1$, and which are advected with the background flow. 
The perturbed fluid moves towards regions of low-entropy,
which generally coincide with underpressures. The entropy-vortex perturbation ($\delta K$) has opposite
polarity than $\delta S$. When combined $\pi/2$ out-of-phase, a spiral pattern is created
in $\delta S$, $\delta K$, and $\delta v_\phi$. The pressure, however, has
a different behavior. 
In a sloshing mode, it propagates outwards, as can be seen in the animated version of 
Figure~\ref{f:figure_spiral}. The latter suggests that 
pressure perturbations arise in response to negative entropy perturbations, which 
affect the cooling and thus hydrostatic equilibrium (as in the case of $\ell=0$ modes, \citealt{FT09a}). 
The source terms in the
differential system (eqns.~[\ref{eq:h_evol}]-[\ref{eq:K_evol}]) are such that cooling terms are 
comparable to terms proportional to $\omega$ for fundamental modes (which have the lowest 
$|\omega|$ for a given $\ell$), 
decreasing their relative contribution for higher overtones as $\omega$ increases in magnitude.
When combined $\pi/2$ out-of-phase, the pressure perturbations from sloshing modes result 
in two almost half-circular regions of opposite polarity  oriented towards the
center of the prograde and retrograde fluid streams. 
As the amplitude grows, a protuberance and depression
arise at opposite sides of the shock, following the pressure perturbation (\S\ref{s:time_dep}).
The overall flow structure rotates with a period equal to the oscillation frequency of 
the sloshing modes. 

\begin{figure*} 
\includegraphics*[width=0.285\textwidth]{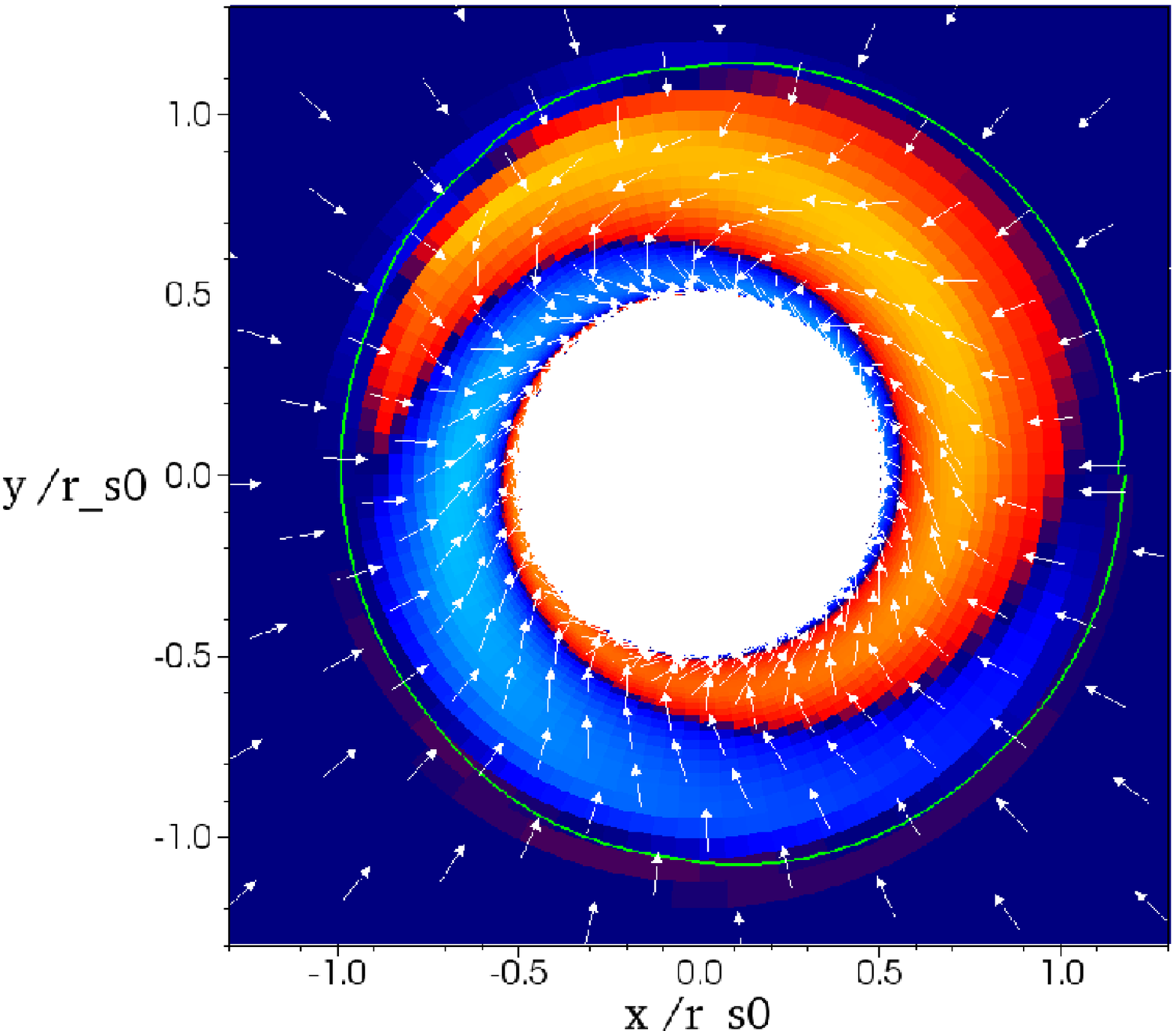}
\includegraphics*[width=0.23\textwidth]{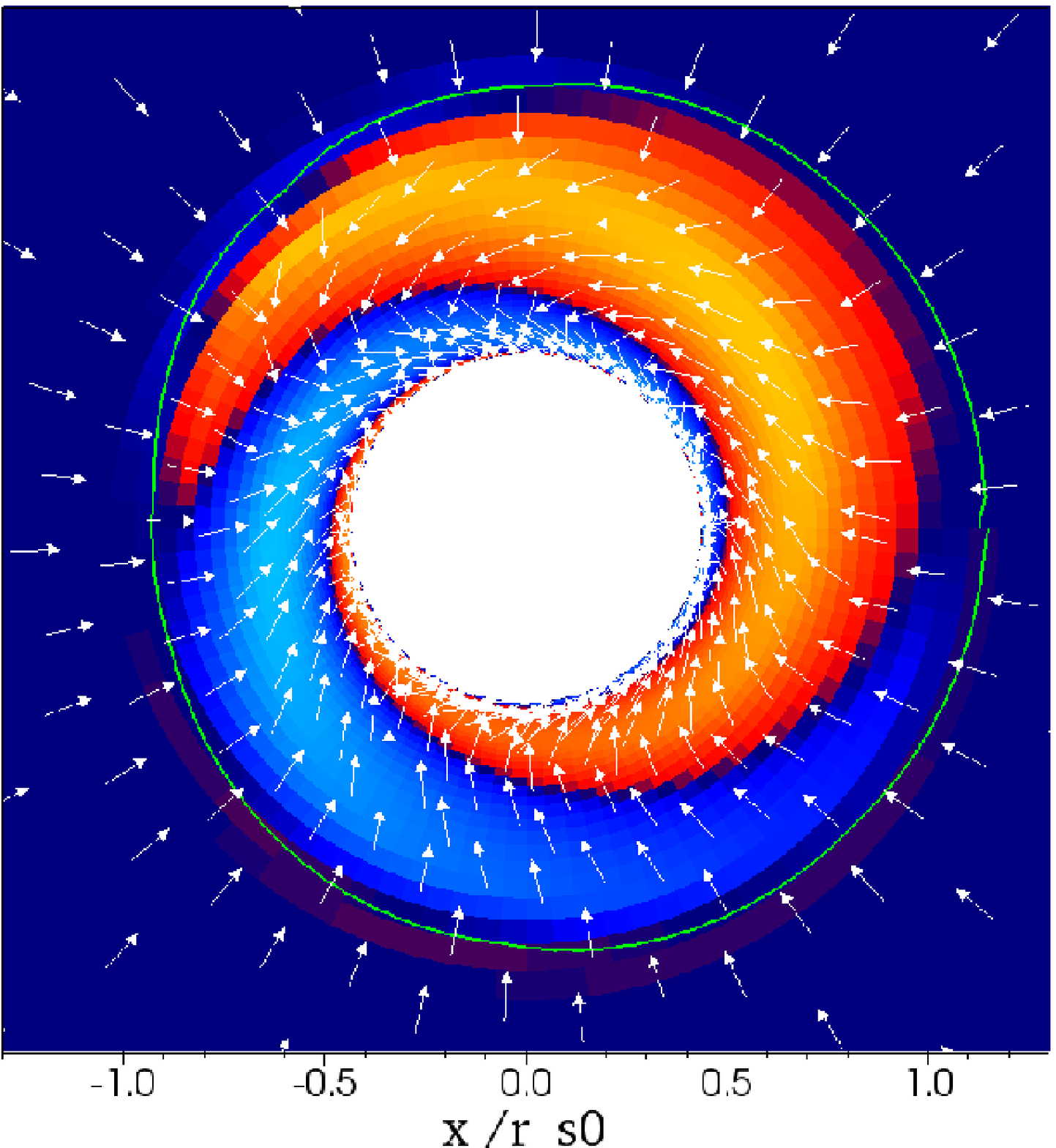}
\includegraphics*[width=0.23\textwidth]{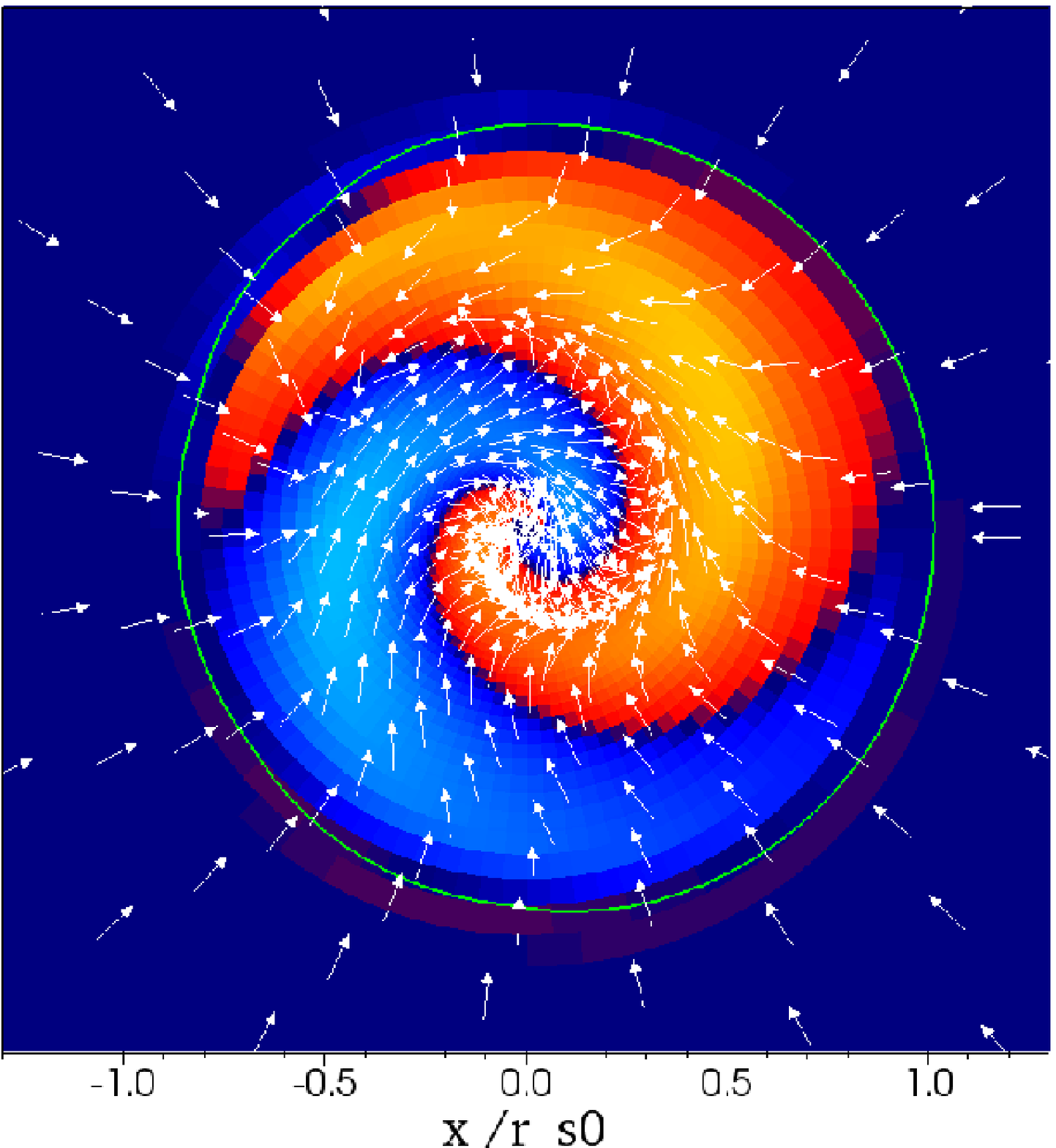}
\includegraphics*[width=0.23\textwidth]{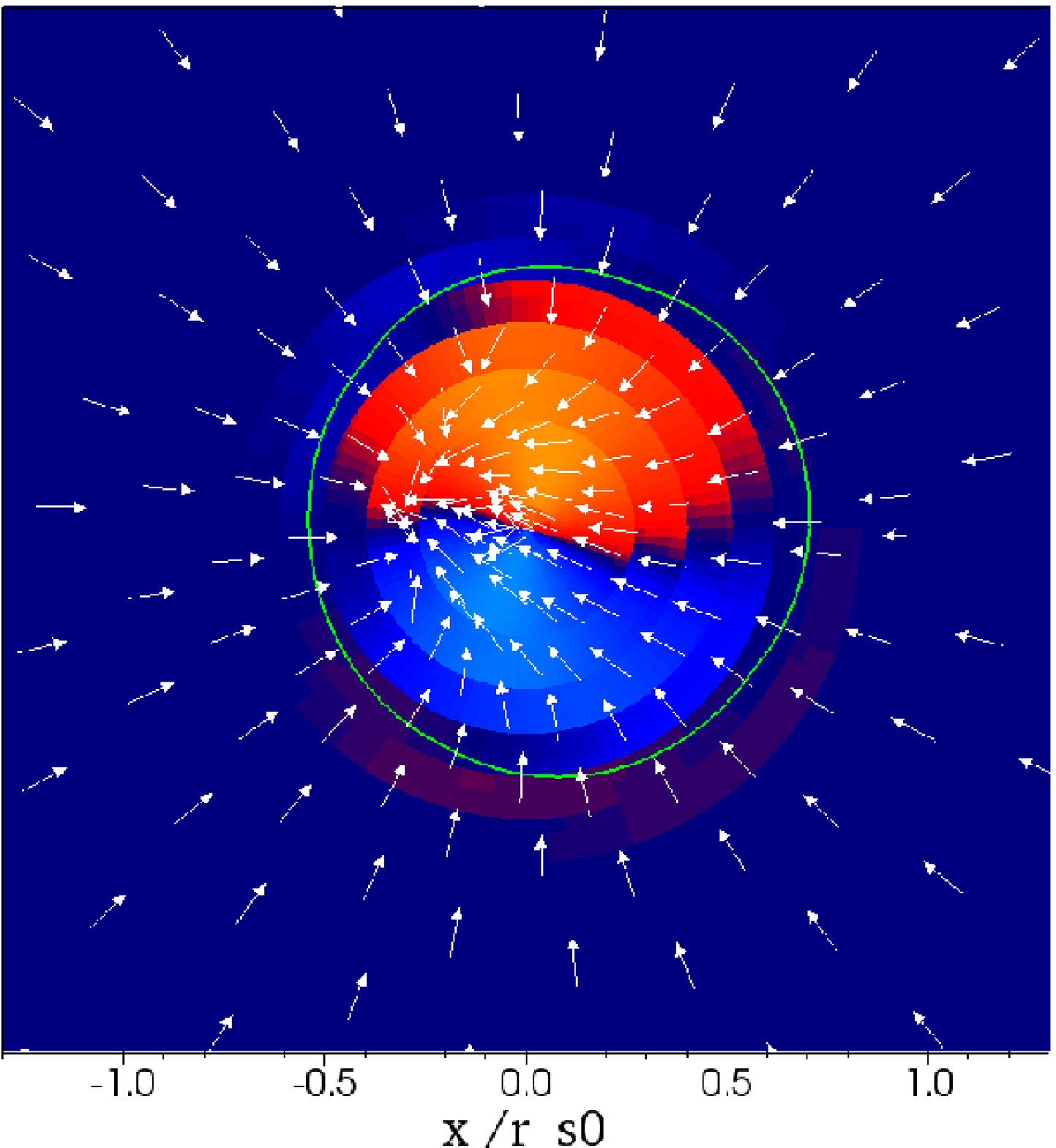}
\caption{Azimuthal velocity in a snapshot of model R5\_L11P2 at time $t = 33\tff0$. 
Units and color table are the same as in Figure~\ref{f:figure_modes},
and panels from left to right correspond to the same altitudes above the equator. The ratio of stellar to
shock radius is $r_*/\rs0 = 0.5$, for which only the fundamental $\ell=1$ mode is unstable.
Despite the relatively coarse grid resolution, the agreement with the linear eigenmodes is excellent.
The green contour shows the surface with internal pressure equal to $0.1\rho_1 v_{\rm ff0}^2$.
An animated version of this figure is available in the online version of the article.
}
\label{f:vphi_zmp}
\end{figure*} 

From the previous discussion, it can be seen that spiral modes need not be limited to the 
x-y plane, a phase shift of $\pi/2$, or to have equal amplitudes.
Panel (a) of Figure~\ref{f:figurepars} shows the effects of decreasing the phase difference from 
$\pi/2$ to $\pi/8$ relative to the standard case shown in panel (l) of Figure~\ref{f:figure_spiral}. The spiral
pattern is still apparent, though not static anymore. Indeed, the animated version of Figure~\ref{f:figurepars}
shows that it is in fact episodic, arising whenever the two sloshing modes interact constructively. 
Panel (b) of Figure~\ref{f:figurepars} shows what happens when the amplitudes are not the
same: again, the spiral pattern is still distinguishable, although more irregular.
We take these results as an indication that a more general class of spiral-like modes
can be obtained by combining sloshing modes in different ways. This means that a larger
region of parameter space (number of modes, relative phase, relative amplitude) of initial
perturbations can result in spiral-like behavior of the shock compared with the 
limiting case of $\pm\pi/2$ phase or equal amplitudes. In \S\ref{s:time_dep} we show that in fact
these modes survive in the nonlinear phase, leading to angular momentum redistribution
below the shock (albeit smaller in magnitude than a phase difference of $\pi/2$).

Adding a third sloshing mode in the z-direction ($Y_1^0$) results in a 45 degree 
tilt of the ``spiral plane" relative to the equator. The azimuthal and polar angles 
determining the normal to this plane are set by the phase and amplitude
of this mode relative to their transverse counterparts, respectively. One can thus immediately
see that the torque imparted to the star by nonlinear spiral modes
is in a direction set by the way
the instability is excited. This conclusion agrees with the results of 
\citet{blondin07a}.  

In the parameter regime relevant to core-collapse supernovae, quadrupolar modes are also significant.
As with the dipolar case, each of these corresponds to an elementary sloshing SASI mode, which can be 
combined and excited in phase
to generate a sloshing mode along an arbitrary axis.
Spiral modes are again generated by exciting at least two of these basic modes out-of-phase. 
Figure~\ref{f:figure_modes} compares the perturbation to the
azimuthal velocity for dipolar and quadrupolar spiral modes at different altitudes from the
equatorial plane. The $|m|=\ell$ modes preserve their spiral structure from pole to pole, while the $|m|<\ell$ 
case has a purely radial flow at the equator, and dipolar spiral flows above and below. These dipolar spirals
are 180 degrees out of phase with each other.

Increasing the angular degree of the mode results in more complicated combinations, although the
modes $Y_\ell^{\pm \ell}$ always have the form $\sin^\ell\theta\, \{\cos(\ell\theta),\sin(\ell\theta)\}$.
This guarantees the existence of two spiral modes with $2\ell$ arms of alternating polarity for all $\ell\geq 1$.

\section{Results from Time-Dependent Simulations}
\label{s:time_dep}

\subsection{Linear Growth and Saturation}
\label{s:time_dep_linear}

In order to trigger an individual spiral mode in our simulations, we drop overdense shells with angular
dependence determined by real spherical harmonics (eqns.~[\ref{eq:Y_z}]-[\ref{eq:Y_2-2}]). 
We locate these shells in the upstream flow in such a way that when advected, they arrive at 
the shock with a time delay that corresponds to a relative phase $\Phi$. In other words, the radial 
positions $r_1$ and $r_2$ of a two-shell combination satisfy
\begin{equation}
\label{eq:trigger_eqn}
\int_{r_1}^{r_2} \frac{\totd r}{|v_r|} = \frac{\Phi}{\omega_{\rm osc}}.
\end{equation}
More than two modes are included straightforwardly, as only relative phases are relevant.
This type of perturbation results in the excitation of an isolated spiral mode
when the background accretion flow does not rotate, allowing comparison with eigenfunctions from 
linear stability analysis, and quantification of its contribution to the angular momentum redistribution
as it grows exponentially in amplitude.

Figure~\ref{f:vphi_zmp} shows the azimuthal velocity field at the same altitudes above the equator as in
Figure~\ref{f:figure_modes} for model R5\_L11P2, for which only the fundamental mode is unstable. 
Despite the coarseness of the grid, the flow structure in the upper row of Figure~\ref{f:figure_modes}
is clearly reproduced\footnote{Despite the fact that the amplitudes of the sloshing modes are slightly 
different.}. The online version of the article contains an animated version of this figure.

\begin{figure}
\includegraphics*[width=0.52\columnwidth]{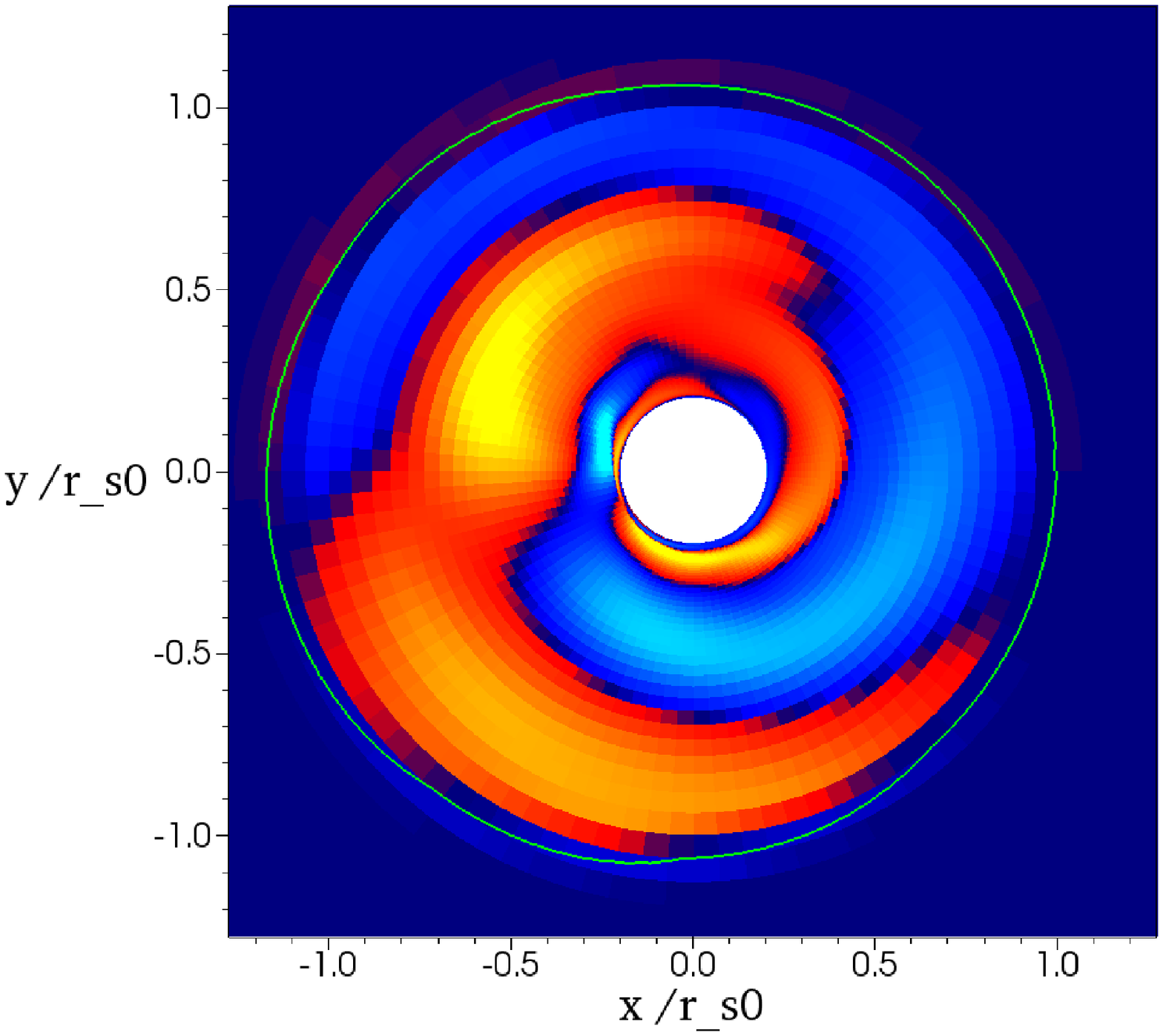}
\includegraphics*[width=0.455\columnwidth]{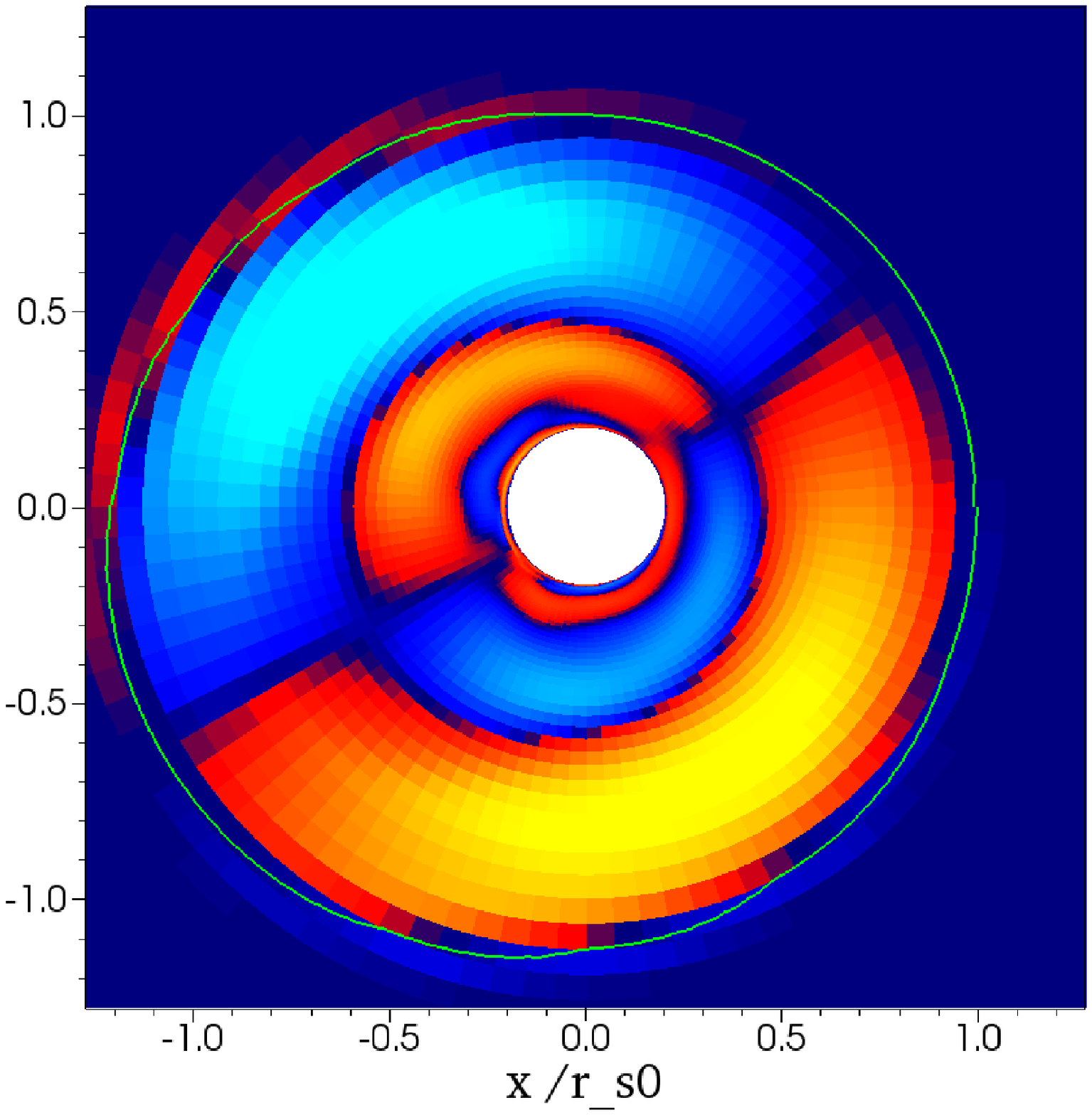}
\caption{Evolution of the azimuthal velocity in the equatorial plane at $t\simeq 30\tff0$ for models
R2\_L11f (left) and R2\_L11h (right), which are closer in size to stalled supernova shocks, but have 
several unstable $\ell=1$ overtones. When the fundamental is excited (left), an irregular spiral
is obtained, whereas an excitation of the first overtone yields
an oblique sloshing mode at the fundamental frequency (right). 
The color scale is the same as in Figure~\ref{f:vphi_zmp}.
}
\label{f:slice_r0.2}
\end{figure}

To test the effects of unstable overtones, we have also evolved a configuration with 
$r_*/\rs0 = 0.2$, which is closer in size to stalled supernova shocks (models R2\_L11f and R2\_L11h). 
In this case $\ell=1$ has
several unstable overtones (see, e.g., Figure~13 of \citealt{FT09a}), which are all excited
whenever an $\ell=1$ perturbation is imposed. Figure~\ref{f:slice_r0.2} shows
the resulting azimuthal velocity arising from perturbations of the type in equation~(\ref{eq:trigger_eqn}), 
employing phase differences corresponding to the fundamental and first overtones of the $Y_1^1+iY_1^{-1}$ 
mode. In the first case, shown in the left panel, a spiral-like pattern is formed, although the presence of
unstable overtones makes it irregular. In the second, on the right, an initial first overtone spiral becomes 
overtaken by a fundamental sloshing mode in an oblique direction, because the latter mode 
has a larger growth rate.

\begin{figure*}
\includegraphics*[width=0.32\textwidth]{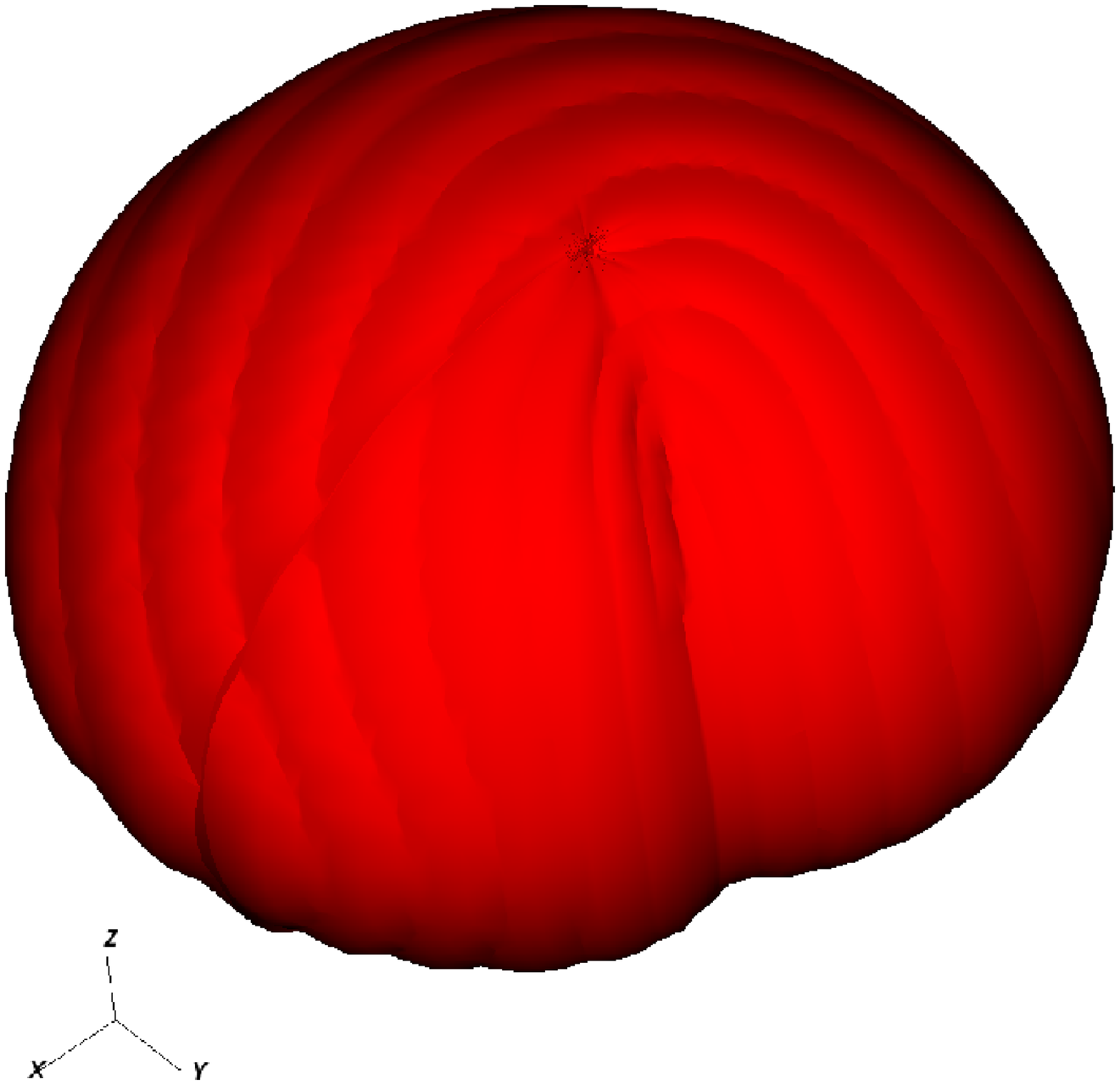}
\includegraphics*[width=0.32\textwidth]{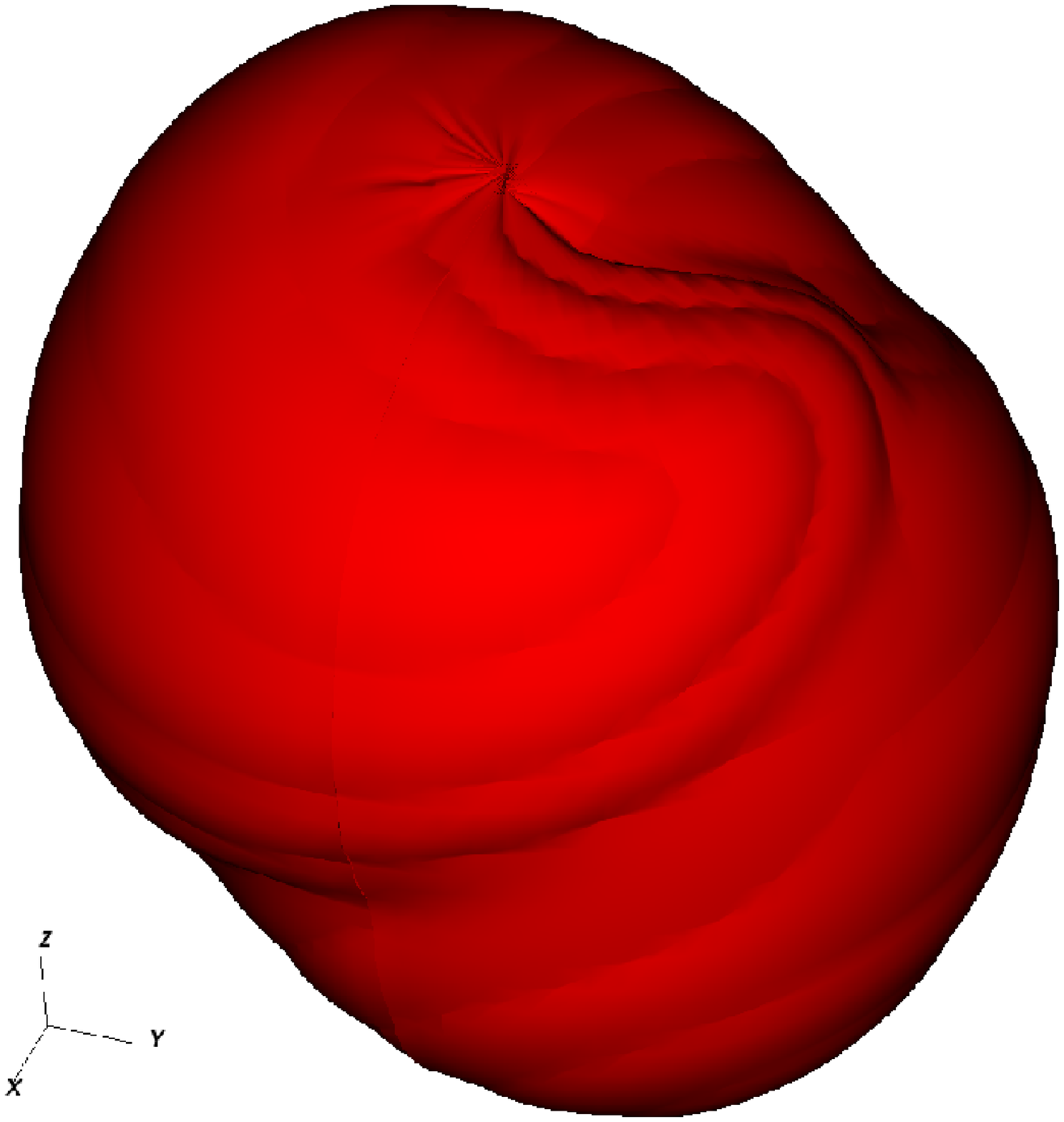}
\includegraphics*[width=0.32\textwidth]{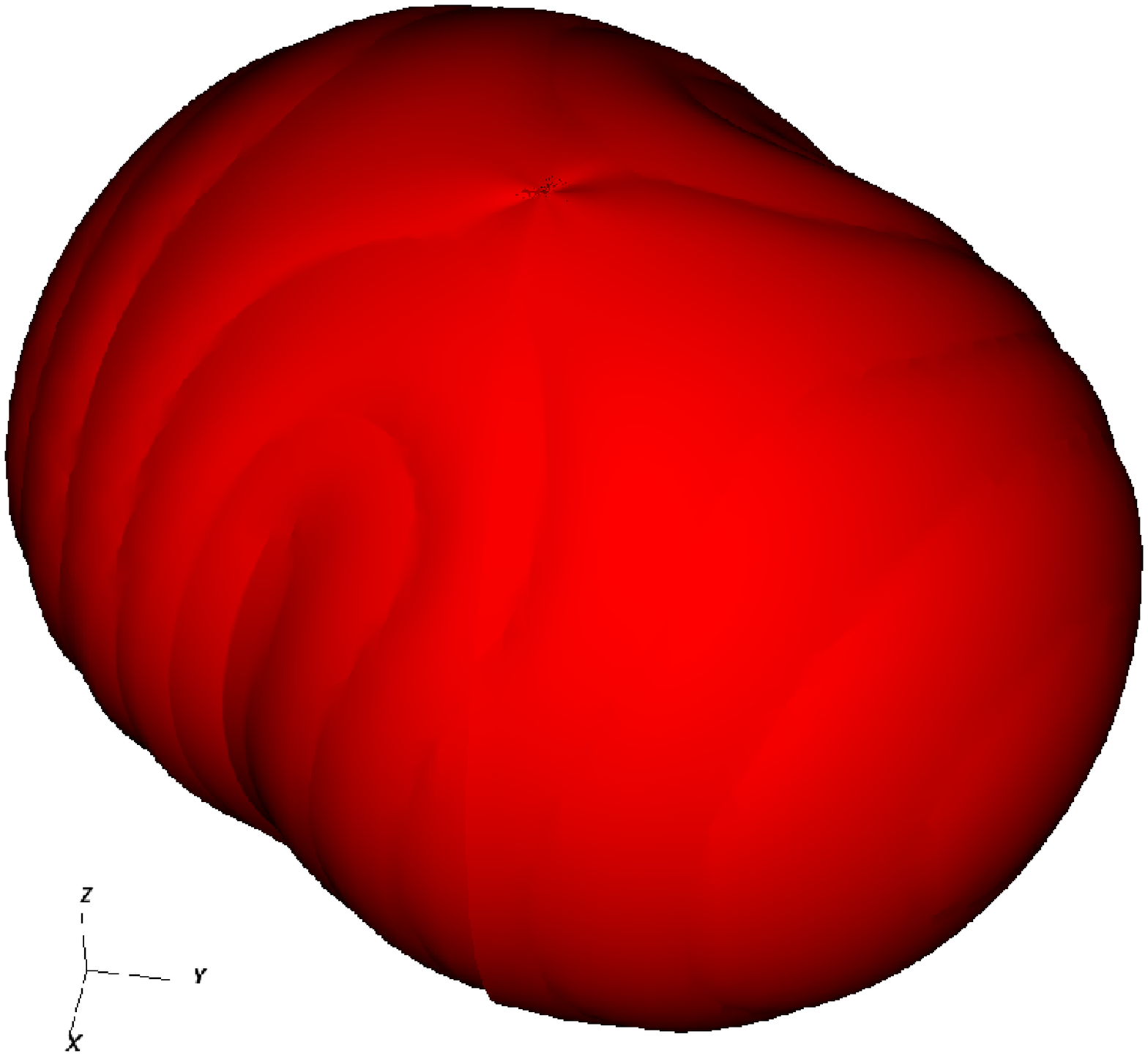}
\caption{Isobaric surfaces ($p = 0.1\rho_1\vff2$), showing the morphology
of the shock in the transition to the nonlinear regime for models
R5\_L11P2 at $t=46.5\tff0$ (left), R6\_L21P2 at $t=57\tff0$ (middle), and R6\_L22P2 at $t=52.5\tff0$ (right).
In all cases, the ratio $r_*/\rs0$ is chosen so that only the fundamental mode is unstable, for each $\ell$.
The meridional line at $y=0$ is an artifact of the plotting tool, while the irregularity
at the north pole is due to the numerical instability described in section~\ref{sec:time_dep_setup}.
In all cases, the shock deformations grow in amplitude and rotate counterclockwise.
An animated version of this Figure is available in the online version of the article.
}
\label{f:contours_zmp}
\end{figure*}

As a mode grows in amplitude, the pressure perturbation shown in Figure~\ref{f:figure_spiral}k
develops into a protuberance that rotates with the oscillation period of the instability. Eventually
a shock forms at the interface where the two streams move towards each other, resulting in a triple
point at the shock \citep{blondin07b}.
Figure~\ref{f:contours_zmp} shows the morphology of the shock right before this triple point forms, 
for models R5\_L11P2, R6\_L21P2, and R6\_L22P2. 

\begin{figure}
\includegraphics*[width=\columnwidth]{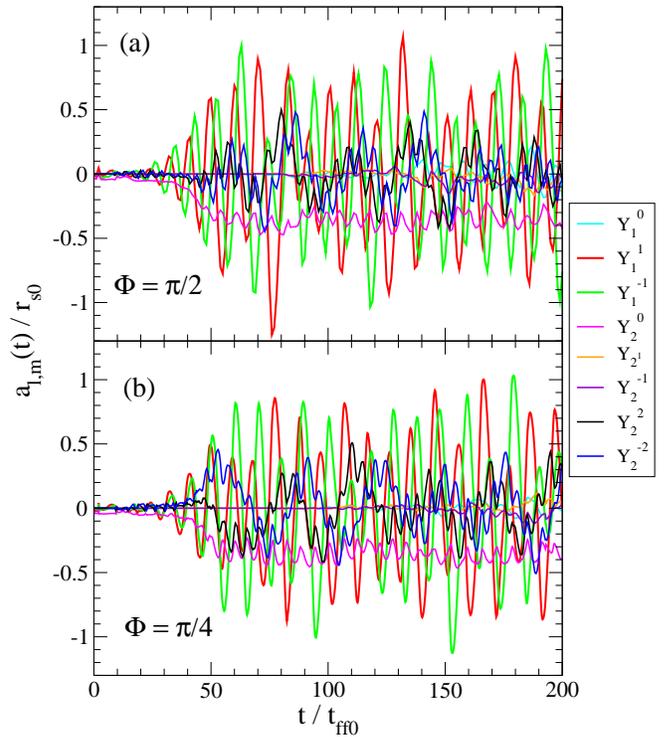}
\caption{Real spherical harmonic coefficients of the shock (traced by an isobaric surface)
as a function of time, for models R5\_L11P2c (top) and R5\_L11P4 (bottom).
The linear phase lasts approximately until the $\ell=1$ displacement is $\sim 10\%$ of
the postshock cavity, or $t\simeq 30\tff0$. Full saturation is reached around $t=60\tff0$.
Note that the $Y_1^{\pm1}$ coefficients do not fall into the same envelope in the linear
phase, indicating that they were not excited with the same amplitude.
}
\label{f:alm_0_200}
\end{figure}

The initial phase of exponential growth of isolated modes is followed by nonlinear mode
coupling and saturation. Figures \ref{f:alm_0_200}a and \ref{f:alm_0_200}b show the spherical harmonic coefficients
\begin{equation}
\label{eq:alm}
a_{\ell,m}(t) = \int \totd\Omega\, Y_\ell^m\, R_s(\theta,\phi,t),
\end{equation} 
where $R_s(\theta,\phi,t)$ is an isobaric surface tracing the shock, as a function of time 
for models R5\_L11P2c and R5\_L11P4. In both cases, the primary sloshing mode
combination remains isolated in the linear phase, along with a non-oscillatory increase in
the $\ell=2$, $m=0$ mode due to the increasing oblateness of the shock, likely caused by
the redistribution of angular momentum (\S\ref{sec:angular_mom}). Other modes
are excited only after the primary mode reaches nonlinear amplitudes. Interestingly,
the first modes to couple are $Y_2^{\pm2}$, which achieve amplitudes about 50\%
lower than the primary mode. The axisymmetric $\ell=1$ mode together with
$\ell=2$, $m=\{0,\pm1\}$ modes couple much later, and do not reach as large an amplitude as $m=2$.
We do not show $\ell=3$ results to keep Figure~\ref{f:alm_0_200} legible, but a similar trend is observed in that
$Y_3^{\pm 3}$ couple first and reach the largest amplitudes of all $\ell=3$ modes.
This sequence of mode coupling differs from that found by \citet{iwakami08}, who witnessed
the $\ell=2$, $m=0,\pm1$ modes to couple first and reach the largest amplitude behind $\ell=1$.
However, their non-axisymmetric perturbation excites the real spherical harmonics $Y_1^0$ and $Y_1^1$
in phase, hence it is not surprising that the mode coupling sequence differs.

\begin{figure}
\includegraphics*[width=\columnwidth]{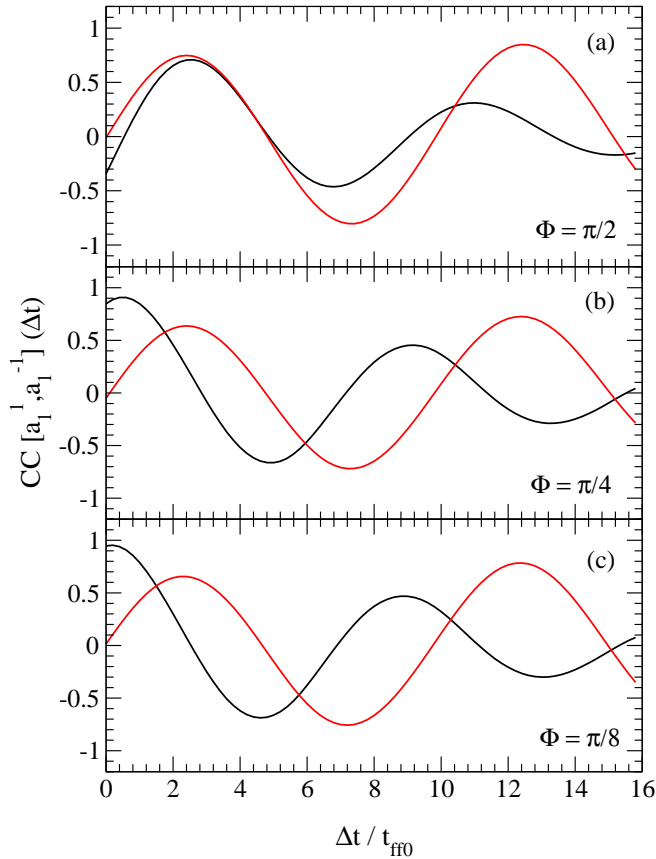}  
\caption{Normalized cross-correlation (eq.~[\ref{eq:cross-corr}]) applied to models R5\_L11P2c (top),
R5\_L11P4 (middle), and R5\_L11P8 (bottom).
Shown is the linear and weakly nonlinear phase (black), covering
the time interval $t \in [24,50]\tff0$, and the fully saturated phase (red), over $t \in [100,200]\tff0$
(see Figure~\ref{f:alm_0_200}). The ratio of consecutive peaks in the linear phase is approximately 
$\exp{(\pi\omega_{\rm grow}/\omega_{\rm osc})}$, or the fraction of an e-folding corresponding to one-half of an 
oscillation period.
The initial condition is reflected in the cross-correlation taken over the linear phase. However, the
time-delay between sloshing modes in the nonlinear phase is almost independent of initial conditions.
}
\label{f:cross-corr}
\end{figure}

To track how the relative phase in the primary spiral mode evolves going into the
nonlinear phase, we compute a normalized cross-correlation between the spherical harmonic
coefficients of the corresponding sloshing modes:
\begin{eqnarray}
\label{eq:cross-corr}
&& \mathrm{CC}[a_{1,1},a_{1,-1}](\Delta t) = \nonumber \\&&\frac{\int\,dt\,\left[a_{1,1}(t)-\bar{a}_{1,1}\right] 
\left[a_{1,-1}(t+\Delta t)-\bar{a}_{1,-1}\right]}
{\left(\int dt\left[a_{1,1}(t)-\bar{a}_{1,1}\right]^2\right)^{1/2}\left(\int\,dt\left[a_{1,-1}(t)-\bar{a}_{1,-1}\right]^2\right)^{1/2}},
\end{eqnarray}
where $\bar{a}_{\ell,m}$ is the time average of $a_{\ell,m}$ and the integrals are performed over the same  
time interval $[t_{\rm{min}},t_{\rm{max}}]$. Figure~\ref{f:cross-corr} shows
the results of applying equation~(\ref{eq:cross-corr}) to the linear and nonlinear phase of modes
R5\_L11P2c, R5\_L11P4, and R5\_L11P4, which differ only in the initial relative phase of the perturbation. 
In the linear regime, the cross-correlation has an initial peak at a time delay $\Delta t_1 = \Phi/\omega_{osc}$. As this 
amounts to shifting the second sloshing mode back into the first, equation~(\ref{eq:cross-corr}) yields values
near unity\footnote{The fact that it is not exactly unity means that both sloshing modes were not excited
with the same amplitude, an effect we believe is due to the partial dissipation of the second
overdense shell when advected by the supersonic flow before crossing the shock.}. 
Subsequent peaks arise at $\Delta t_n = (\Phi + n\pi)/\omega_{osc}$, ($n=1,2,...$), with an amplitude 
$(-1)^n\exp(-n\pi\omega_{\rm grow}/\omega_{\rm osc})$, i.e., each half period down in magnitude by 
the inverse exponential growth rate, with alternating sign. When equation~(\ref{eq:cross-corr})
is applied to the fully saturated nonlinear phase, the first peak shifts to a larger value of $\Delta t$
relative to the linear phase, with no significant change in amplitude between subsequent peaks. 
Furthermore, the three nonlinear cross-correlations are nearly identical, with a period $\simeq 12\tff0$
and a time delay $\simeq 2.5\tff0$, corresponding to a phase shift $\simeq 2\pi/5$. This suggests 
that the development of spiral modes in the fully saturated stage is independent of the initial conditions,
for any non-zero initial relative phase between sloshing modes.
As mentioned in \S\ref{sec:eigenmode}, a larger class of initial perturbations can therefore
result in spiral modes, not just those tuned to a relative phase of $\pi/2$. 

\begin{figure}
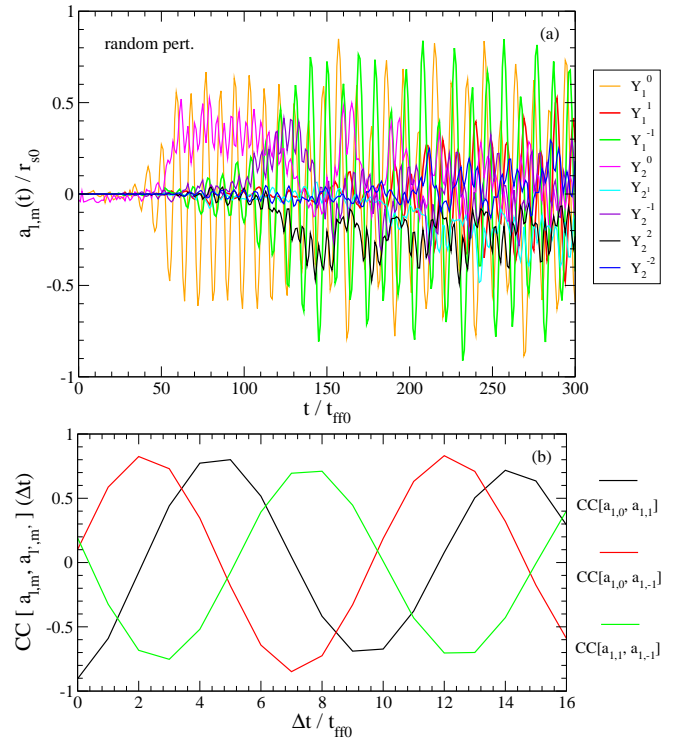

\includegraphics*[width=\columnwidth]{f9a.eps}
\includegraphics*[width=\columnwidth]{f9b.eps}
\caption{\emph{Top}: spherical harmonic coefficients for model R5\_RAN. The initial bias towards axisymmetric ($Y_1^0$ and $Y_2^0$)
perturbations is due to the form of the grid (more cells closer to the poles). Non-axisymmetric modes
do indeed grow out of numerical noise, as expected. \emph{Bottom}: normalized cross-correlation 
(eq.~[\ref{eq:cross-corr}])
applied over the time interval $t \in [200,300]\tff0$ to the $\ell=1$ modes pairwise. All three of them are out-of-phase
with each other in the nonlinear phase, leading to angular momentum redistribution (Figure~\ref{f:angmom_encl_rand}).
}
\label{f:PE}
\end{figure}

We have also performed a simulation in which, instead of using overdense shells, we impose random
cell-to-cell pressure perturbations below the shock (R5\_RAN). For this, we employ the setup with $r_*/\rs0 =0.5$,
for which only the fundamental $\ell=1$ mode is unstable (and the most unstable for all $\ell$). Figure~\ref{f:PE}a
shows the corresponding spherical harmonic coefficients as a function of time. Initially, a sloshing
mode along the $z$-axis is triggered due to the form of the grid. However, as this mode reaches the
nonlinear phase, a sloshing mode along the $y$ axis emerges out-of-phase, and then another along
the $x$ axis follows. Figure~\ref{f:PE}b shows the normalized cross-correlation in the nonlinear phase, for the three
different $\ell=1$ mode combinations over the time interval $t \in [200,300]\tff0$, showing that all of them are out-of phase. 
Note that the phase delay between $Y_1^1$ and $Y_1^{-1}$ is
nearly the same as that in Figure~\ref{f:cross-corr} (modulo half a period), even though $Y_1^1$ is still growing in
amplitude.
The resulting angular momentum
redistribution is discussed in \S\ref{sec:angular_mom}.

To close this subsection, we address the effects of dimensionality and resolution on the saturation
amplitude. Table~\ref{t:amplitudes} shows the rms fluctuation of the $a_{1,1}$ spherical harmonic coefficient around
its mean for various models. There does not appear to be a systematic difference between the saturation
amplitude in 2D and 3D at this resolution and with this numerical method. The changes due to the presence of the cutout around
the polar axis or a resolution doubling are of the same magnitude, $\sim 10\%$. Similarly, whether the sloshing mode is isolated or
part of a spiral mode seems to make a small difference, which is a weak function of the relative phase between modes.
In the context of parasitic instabilities \citep{guilet09a}, this indicates that either (i) the resolution is
still too low to expose the different growth rates of parasites in 2D and 3D, or (ii) given the structure of these
modes, the difference in the growth rates does not translate in sizable differences in the saturation
amplitudes.
\begin{deluxetable}{lclc}
\tablecaption{Saturation Amplitude of $Y_1^1$ mode\label{t:amplitudes}}
\tablewidth{0pt}
\tablehead{
\colhead{Model} &
\colhead{Dimensionality} &
\colhead{Resolution\tablenotemark{a}} &
\colhead{$\Delta a_{1,1}\tablenotemark{b}$}
}
\startdata
r5\_L1\_LR    & 2D &  $\phantom{1}56\times 48$             & 0.505 \\
r5\_L1\_LRc   & 2D &  $\phantom{1}56\times 48$c            & 0.424 \\
r5\_L1\_HR    & 2D &  $          112\times 96$             & 0.412 \\
r5\_L1\_HRc   & 2D &  $          112\times 96$c            & 0.454 \\
R5\_L11x      & 3D &  $\phantom{1}56\times 48\times 96$c   & 0.473 \\
R5\_L11P2     & 3D &  $\phantom{1}56\times 48\times 96$    & 0.505 \\
R5\_L11P2c    & 3D &  $\phantom{1}56\times 48\times 96$c   & 0.499 \\
R5\_L11P4    & 3D &  $\phantom{1}56\times 48\times 96$c    & 0.504 \\
R5\_L11P8    & 3D &  $\phantom{1}56\times 48\times 96$c    & 0.481 \\
R5\_L11\_HR   & 3D &  $          112\times 96\times 192$c  & 0.525
\enddata
\tablenotetext{a}{The letter c denotes a 5 degree cutout around the polar axis.}
\tablenotetext{b}{For models R5\_L11P2 and R5\_L11\_HR the time range employed
is $[100,147]\tff0$ and $[100,180]\tff0$, respectively. For all others, it is $[100,200]\tff0$.}
\end{deluxetable}

\subsection{Angular Momentum Redistribution}
\label{sec:angular_mom}

Given that the background accretion flow has not net angular momentum, any net spin-up
of the protoneutron star via accreted matter involves the separation of the postshock
flow into regions with angular momentum of opposite sign (e.g., \citealt{blondin07b}). 
In what follows, we explore how the linear and nonlinear phases of spiral modes mediate 
this angular momentum redistribution, and quantify the magnitude of the effect. To ease comparison
with other studies, we express angular momenta in units of $\dot{M}\rs0^2$, which amounts to
dividing our results by $4\pi$ within the dimensionless unit system  described in \S\ref{sec:phys_model}.

\subsubsection{Linear Phase} 
\label{sec:linear_spinup}

\begin{figure}
\includegraphics*[width=\columnwidth]{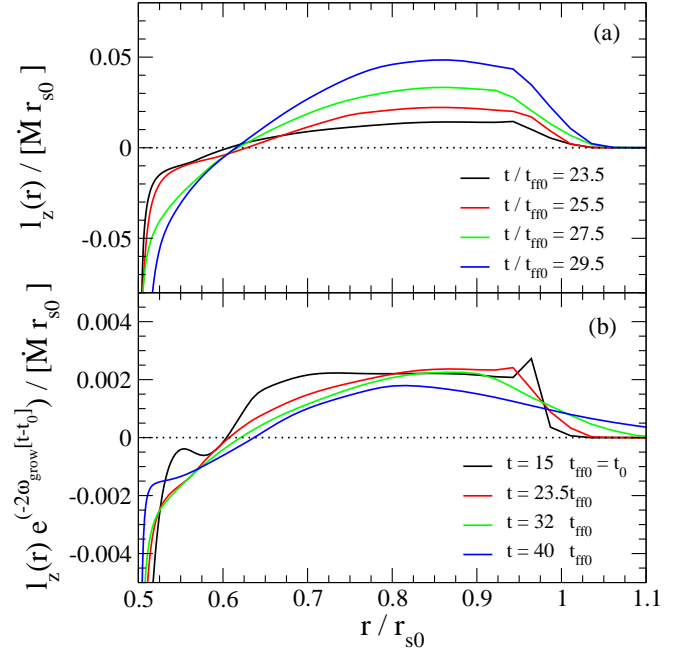}
\caption{Z-component of the angular momentum density integrated over a spherical surface $l_z(r)$ (eq.~[\ref{eq:angular_mom_density}])
as a function of time for model R5\_L11\_HR. \emph{Top:} Values at four times covering an entire $a_{1,1}$ oscillation period,
showing that $l_z(r)$ is non-oscillatory. \emph{Bottom:} values at a few times where the $Y_1^1$ mode is at phase zero, scaled by
the square of the mode amplification factor relative to $t_0 = 15\tff0$, $\exp(-2\omega_{\rm grow}[t-t_0])$, showing that $l_z(r)$
grows at nearly twice the growth rate of the primary linear spiral mode.
}
\label{f:lz_time}
\end{figure}

Figure~\ref{f:lz_time} shows the z-component of the angular momentum density integrated over a
spherical surface at radius $r$
\begin{equation}
\label{eq:angular_mom_density}
l_z(r) = r^2\int\, \totd\Omega\, \rho\, r\sin\theta\, v_\phi,
\end{equation}
for model R5\_L11\_HR at various times during the linear phase. The upper panel shows
times covering a full oscillation cycle, and the bottom shows curves rescaled by a 
factor $\exp{[-2\omega_{\rm grow}(t-t_0)]}$ over a longer timescale, with $\omega_{\rm grow}$ the growth
rate measured from the spherical harmonic coefficients. We thus infer that,
in the linear phase,
(i) a spiral mode has a characteristic angular momentum density profile
      which separates the flow into two counter-rotating regions,
(ii) this profile is non-oscillatory, and
(iii) it grows in amplitude at nearly twice the growth rate of the mode.
The latter is consistent with the fact that $l_z(r)$ is a second-order quantity, 
as the zeroth- and first-order components of equation~(\ref{eq:angular_mom_density}) vanish.
To lowest order in $\Delta \xi/\rs0$, it is made up of products of the form $\Delta\rho^{(1)}\Delta v_\phi^{(1)}$ and 
$\rho^{(0)}\Delta v_\phi^{(2)}$, where the superscripts in parentheses indicate explicitly the order
of the perturbation.

We can separate out equation~(\ref{eq:angular_mom_density}) into components
\begin{eqnarray}
\label{eq:ang_mom_dens_components}
l_z(r)     & = & l^{(0)(2)} + l^{(1)(1)} + {\rm O(3)} \\
\label{eq:l_02_def}
l^{(0)(2)} & = & r^3\int\,\totd\Omega\, \rho^{(0)}\, \sin\theta\Delta v_\phi^{(2)} \\
\label{eq:l_11_def}
l^{(1)(1)} & = & r^3\int\,\totd\Omega\, \Delta \rho^{(1)}\,\sin\theta\Delta v_\phi^{(1)}
\end{eqnarray}
where O(3) denotes higher order terms. Due to the fact that $\rho^{(0)}$ is spherically symmetric, 
only the $\ell=0$ component of $\sin\theta\,\Delta v_\phi^{(2)}$ survives:
\begin{equation}
\label{eq:v_phi_second_order}
\left[\sin\theta\Delta v_{\phi}^{(2)}\right]_{0,0} = \int\,\totd\Omega\, Y_0^0\,\sin\theta\,v_\phi - {\rm O(3)}.
\end{equation}
The first order components are the real parts of the complex perturbations
\begin{eqnarray}
\Delta \rho^{(1)}             & = & \frac{1}{2}\sum_{\ell,m}\left[ \delta \tilde{\rho}_{\ell,m}\Upsilon_\ell^m e^{-i\omega_{\ell,m} t} +
                                                         \delta \tilde{\rho}^*_{\ell,m}\Upsilon_\ell^{m*} e^{i\omega^*_{\ell,m}t}\right]\\
\sin\theta\Delta v_\phi^{(1)} & = & \frac{i}{2}\sum_{\ell,m}m\left[ \delta \tilde{v}_{\Omega,\ell,m}\Upsilon_\ell^m e^{-i\omega_{\ell,m} t}\right.
                                     \nonumber\\ 
                              &   & \qquad\qquad - \left.\delta \tilde{v}^*_{\Omega\ell,m}\Upsilon_\ell^{m*} e^{i\omega^*_{\ell,m}t}\right],
\end{eqnarray}
where $\Upsilon_\ell^m$ denotes the usual complex spherical harmonics, and the star stands for complex conjugation. 
Integrating over angles yields
\begin{eqnarray}
\label{eq:I11}
l^{(1)(1)}  = r^3\sum_{\ell,m} m\, & & e^{2{\rm Im}(\omega_{\ell,m})t}\, \left[ {\rm Im}(\delta\tilde{\rho}_{\ell,m}){\rm Re}(\delta\tilde{v}_{\Omega,\ell,m})\nonumber\right.\\
           & &\left.  - {\rm Re}(\delta\tilde{\rho}_{\ell,m}){\rm Im}(\delta\tilde{v}_{\Omega,\ell,m})\right] +{\rm O(3)}.
\end{eqnarray}

\begin{figure}
\includegraphics*[width=\columnwidth]{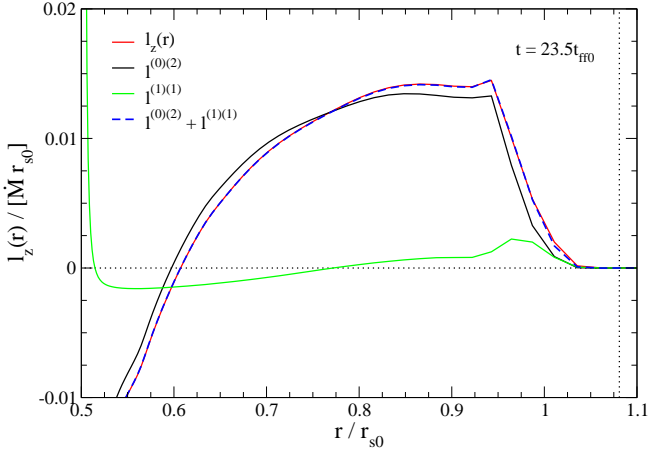}
\caption{Angular momentum density integrated over a spherical surface for model R5\_L11\_HR (red line, 
eq.~[\ref{eq:angular_mom_density}]) as a function of radius, evaluated at $t=23.5\tff0$, 
where the spiral mode is at phase zero. Also shown are the dominant second order components $l^{(0)(2)}$
and $l^{(1)(1)}$ (defined in eqns.~[\ref{eq:ang_mom_dens_components}]-[\ref{eq:l_11_def}], 
black and green lines, respectively),
where the latter is evaluated for the $\ell=1$, $m=1$ mode only (eq.~[\ref{eq:I11}]), and their sum 
(blue dashed lines). The vertical dotted line shows the average shock radius. 
}
\label{f:angmom_components}
\end{figure}

Figure~\ref{f:angmom_components} shows $l_z(r)$ at time $t=23.5\tff0$ for model R5\_L11\_HR, where the $Y_1^1+iY_1^{-1}$ 
mode is at phase zero (cf. Figure~\ref{f:alm_0_200}). Also shown
are $l^{(0)(2)}$, with both $\rho^{(0)}$ and $\sin\theta\,\Delta v_\phi^{(2)}$ obtained by projection of the corresponding fields
fields onto $Y_0^0$, and the $\ell=1$, $m=1$ component of $l^{(1)(1)}$. For the latter, the real part of the complex
amplitude at phase zero is the projection onto $Y_1^1$, while the imaginary part is 
minus the $Y_1^{-1}$ component. Most of the angular momentum density comes from $l^{(0)(2)}$, 
with $l^{(1)(1)}$ being a $\sim 10\%$ correction. Together, these terms account for almost all of $l_z(r)$, in agreement
with the fact that modes with $\ell\ge 2$, $m\ne 0$ have a negligible amplitude in the linear phase. 

\begin{figure}
\includegraphics*[width=\columnwidth]{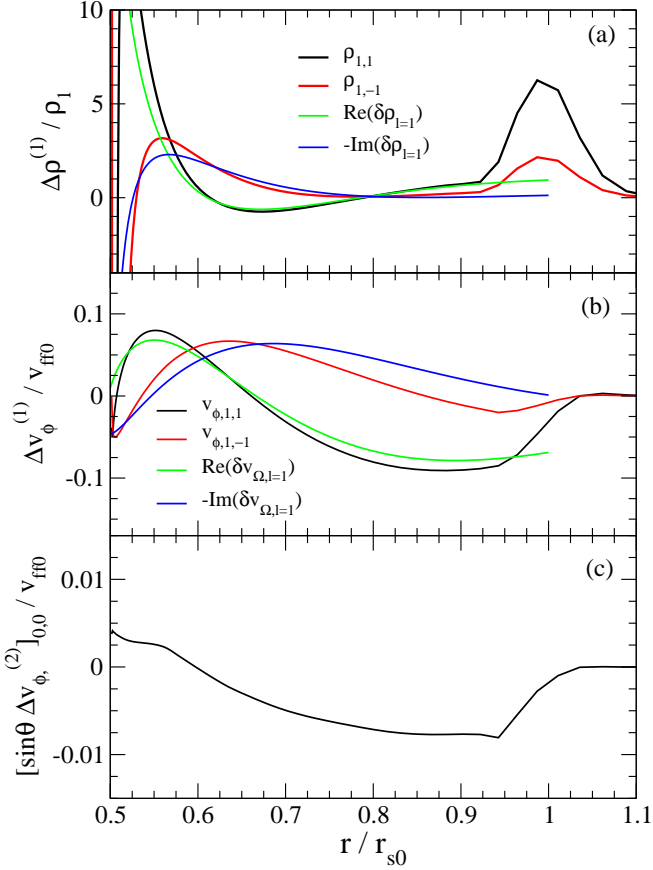}    
\caption{\emph{Top:} $\ell=1$, $m=\pm 1$ components of the density at time $t=23.5\tff0$ for model R5\_L11\_HR (black
and red curves). Also shown are the corresponding values predicted by linear stability theory for a relative phase $\Phi=\pi/2$,
with $Y_1^1$ at zero phase. The amplitude of the eigenfunctions has been normalized by the value of the shock spherical harmonic
coefficients. \emph{Middle:} same as top, but now for the azimuthal velocity (see Appendix~\ref{sec:v_trans} for the definition
of $\delta v_\Omega$). The reasonable agreement with eigenfunctions means that $l^{(1)(1)}$ is mostly accounted for by
linear theory. \emph{Bottom:} The projection of the azimuthal velocity onto $Y_0^0$, which is second order (eq.~[\ref{eq:v_phi_second_order}]).
Note that its amplitude is about 10 times smaller than the first-order $\ell=1$ perturbations in panel (b).
}
\label{f:angmomdens_eigenfunctions}
\end{figure}

For comparison, figures \ref{f:angmomdens_eigenfunctions}a and b show the $Y_1^{\pm 1}$ components of the density and 
azimuthal velocity fields as a function of radius at $t=23.5\tff0$, for model R5\_L11\_HR, along with the eigenmodes from 
linear stability analysis normalized by the value of their respective spherical harmonic coefficients. Reasonable 
agreement is found between both approaches. Thus, most of $l^{(1)(1)}$ can be accounted for by linear theory. 
Figure~\ref{f:angmomdens_eigenfunctions}c shows
$[\sin\theta\,v_\phi^{(2)}]_{0,0}$ (eq.~[\ref{eq:v_phi_second_order}]). Note that the amplitude is $\sim 10$ times smaller than
the first order $\ell=1$ components in Figure~\ref{f:angmomdens_eigenfunctions}a. 

The angular momentum redistribution is a smooth process, mediated by exponential amplification.
Thus, the formation of a triple point at the shock is not the main agent behind the
the spin-up of the inner region, as we have shown that angular momentum redistribution begins 
as soon as linear modes are excited. The triple point may, however, be related to the saturation of the primary mode, 
setting the magnitude of the angular momentum redistribution.

A key element for a predictive theory of the spin-up due to exponentially growing spiral modes involves
calculation of the second order velocity perturbation $\sin\theta\, v_\phi^{(2)}$, since $l^{(0)(2)}$ dominates the spin-up. 
We surmise that this would require repeating the steps followed for linear perturbations (\ref{s:background}), 
but now expanding to second order in $\Delta \xi/\rs0$. Combining this result with a criterion for the 
saturation of the SASI (such as that from \citealt{guilet09a}) would allow estimation of the maximum angular
momentum redistribution achievable by this means.

\subsubsection{Nonlinear Phase and Total Spin-Up}
\label{sec:nonlinear_spinup}

\begin{figure}
\includegraphics*[width=\columnwidth]{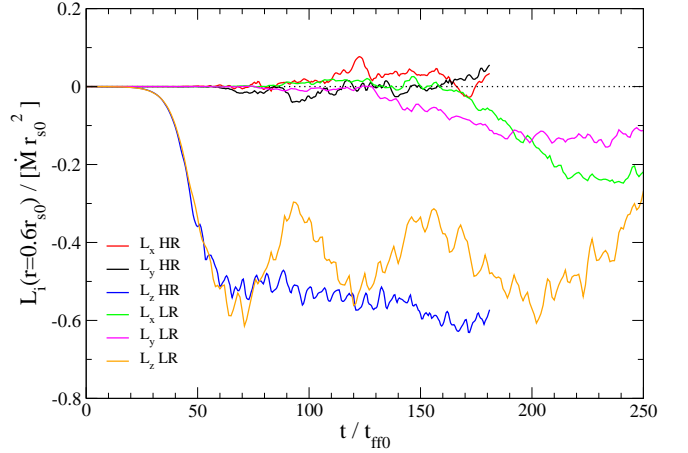}
\caption{Evolution of the three cartesian components of the enclosed angular momentum (eq.~[\ref{eq:angmom_enc}]) evaluated
at $r=0.6\rs0$, where $l_z(r)$ changes sign, for $\ell=1$, $m=\pm 1$ spiral modes at different resolution. Shown 
are $L_x$, $L_y$ and $L_z$ for model R5\_L11\_HR (red, black, and blue curves, respectively) and R5\_L11P2c (green, magenta,
and orange curves, respectively). The bulk of the angular momentum redistribution occurs during the exponentially growing phase,
and the saturation value is a weak function of resolution. The subsequent evolution shows larger departures. Model R5\_L11\_HR
is only evolved until $t=180\tff0$.
}
\label{f:angmom_enc_res}
\end{figure}

We focus first on the high resolution $\ell=1$ model, R5\_L11\_HR, and then present results for other parameter
combinations (Table~\ref{t:models}). 
Figure~\ref{f:angmom_enc_res} shows the evolution of the three components of the
enclosed angular momentum
\begin{equation}
\label{eq:angmom_enc}
L_i(r) = \int_{r_*}^r\, l_i(s)\,\totd s\, \qquad (i=x,y,z)
\end{equation}
evaluated at $r = 0.6\rs0$, 
corresponding to the approximate radius at which $l_z(r)$ changes sign (see Figure~\ref{f:lz_time}). 
Two clear phases can be distinguished. First, from $t=0$ to $50\tff0$, $L_z$ grows exponentially due to the
effects discussed in \S\ref{sec:linear_spinup}. Once other modes start to grow in amplitude, the torque becomes
oscillatory and saturates together with the primary mode. Once full saturation has
been reached, the transverse components $L_x$ and $L_y$ increase in magnitude and fluctuate stochastically
around zero. The subsequent evolution in this fully saturated stage depends somewhat on the resolution employed.
Model R5\_L11\_HR seems to remain in a quasi steady-state until the end of the integration at $t=180\tff0$, while
model R5\_L11P2c shows large amplitude and long period oscillations in $L_z$, and a secular increase
in the magnitude of $L_x$ and $L_y$ at late times.

\begin{figure}
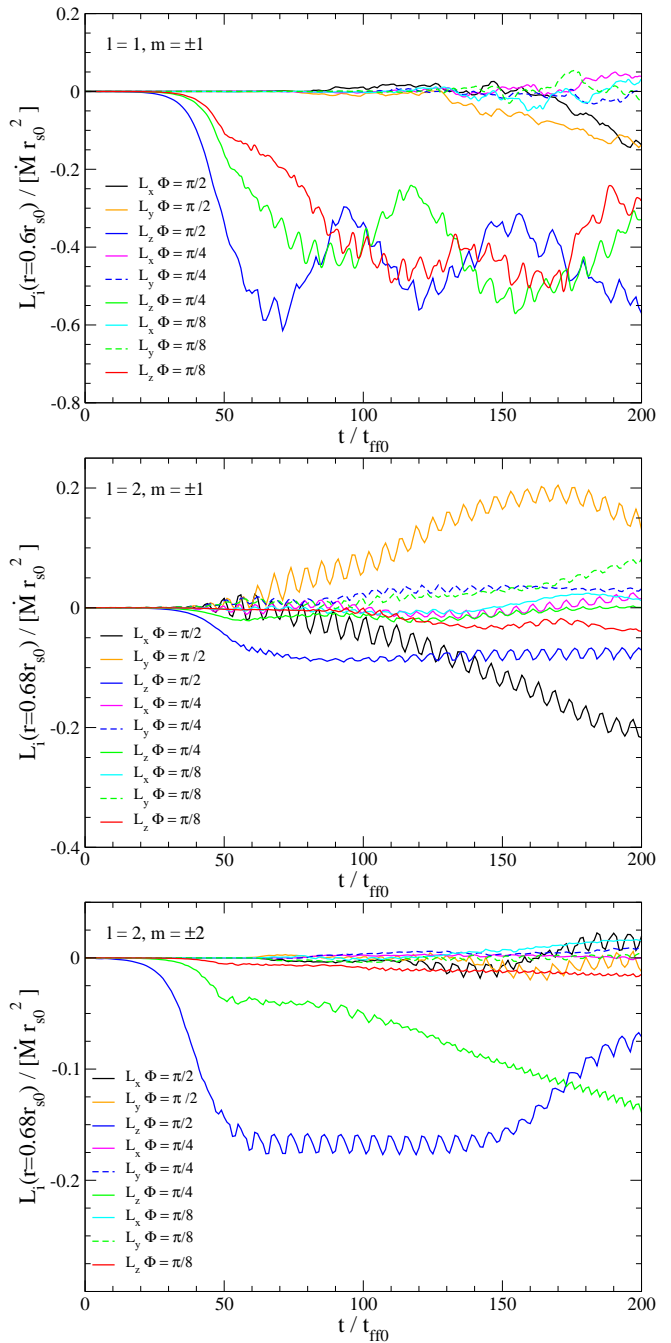

\includegraphics*[width=\columnwidth]{f14a.eps}  
\includegraphics*[width=\columnwidth]{f14b.eps}  
\includegraphics*[width=\columnwidth]{f14c.eps}  
\caption{Enclosed angular momentum components at fiducial radial positions for models where spiral modes with different 
spherical harmonic indices and relative phases are excited. \emph{Top:} models R5\_L11P2c, R5\_L11P4, and R5\_L11P8.
The relative phase of the primary spiral mode is related to the time required to reach maximum spin-up. \emph{Middle:}
Same for models R6\_L21P2c, R6\_L21P4, and R6\_L21P8. This time the angular momentum redistribution is more strongly
dependent on initial relative phase. The radius $r=0.68\rs0$ is equivalent in depth to
that for runs with $r_*/\rs0=0.5$. \emph{Bottom:} Corresponding curves for models R6\_L22P2c, R6\_L22P4, and R6\_L22P8. The behavior
is somewhat similar to the modes with $\ell=1$, $m=\pm 1$, but the maximum spin-up does depend on initial phase. 
\emph{Note that the vertical scale is not the same.}
}
\label{f:angmom_encl_phase}
\end{figure}

Figure~\ref{f:angmom_encl_phase} shows how the angular momentum redistribution depends on the 
polar and azimuthal indices of the primary spiral mode, as well as on the relative phase with 
which it is excited. In this case, the ratio $r_*/\rs0$ is chosen so that only the fundamental mode
is unstable for either $\ell=1$ or $\ell=2$, hence the primary spiral mode grows isolated in the
linear phase. The top panel focuses on the low resolution runs that excite the $\ell=1$, $m=\pm 1$ modes
(R5\_L11P2c, R5\_L11P4, and R5\_L11P8). The maximum spin-up due to the primary spiral mode (along $L_z$) is weakly
dependent on the initial relative phase, and has a magnitude $\sim 0.6\dot{M}\rs0^2$. 
However, the time-delay required to achieve this maximum can differ
by a factor of two. The transverse components ($L_x$ and $L_y$) do not show a significant difference, aside from the
secular growth in the case $\Phi=\pi/2$, which is likely due to resolution effects (Figure~\ref{f:angmom_enc_res}).
Figure~\ref{f:angmom_encl_phase}b focuses on the $\ell=2$, $m=\pm 1$ modes, which still impart a spin-up
along the z-axis despite having vanishing azimuthal velocity at the equator. Note however that the magnitude
of the spin-up is much smaller than $\ell=1$ spiral modes. The magnitude of the transverse components
at late times for $\Phi=\pi/2$ are to be taken with caution, again due to possible resolution effects. Finally, the 
$\ell=2$, $m=\pm 2$ modes show a behavior qualitatively similar to $\ell=1$, $m=\pm 1$, although the magnitude
of the maximum angular momentum redistribution depends strongly on the relative phase. As with $\ell=2$, $m=\pm 1$,
the maximum spin-up is much smaller in magnitude than $\ell=1$.

\begin{figure}
\includegraphics*[width=\columnwidth]{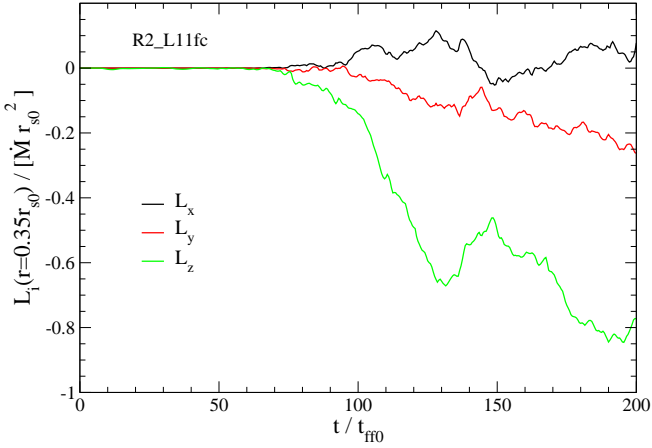}
\caption{Enclosed angular momentum within $r=0.35\rs0$ for model R2\_L11fc, for which $\ell=1$ has several unstable overtones 
(see Figure~\ref{f:slice_r0.2}). The linear phase is qualitatively similar to that in model R5\_L11P2c, but 
the maximum angular momentum is larger at longer times.
}
\label{f:angmom_encl_R0.2}
\end{figure}

The effects of changing the size of the postshock cavity ($r_*/\rs0$) on the angular momentum redistribution are shown
in Figure~\ref{f:angmom_encl_R0.2} for model R2\_L11fc. Here, as in Figure~\ref{f:slice_r0.2}a, the system has a size 
more similar to realistic core-collapse
situations, but is such that many $\ell=1$ overtones are unstable. Still, a spin-up is observed, which is in fact larger
than the case $r_*/\rs0 = 0.5$. The secular growth of the spin-up, however, is to be taken with
caution due to the low resolution of the model. The angular momentum redistribution
following the phase of linear growth is still $\simeq 0.6\dot{M}\rs0^2$. This initial peak, however, comes after a long time-delay
of $130\tff0\simeq 400 M_{1.3}^{-1/2}(\rs0/150~{\rm km})^{3/2}$~ms.

\begin{figure}
\includegraphics*[width=\columnwidth]{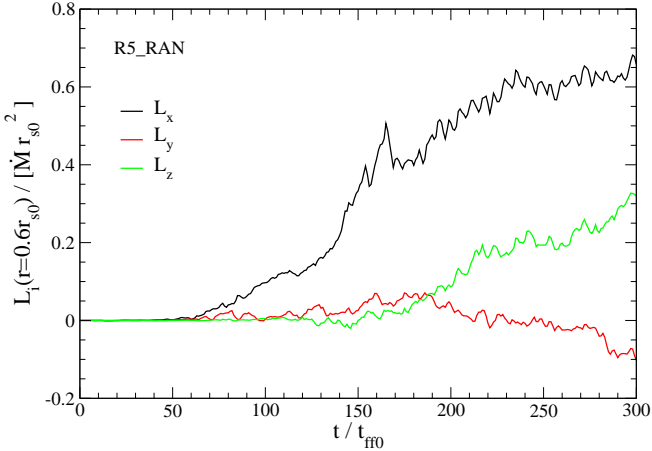}
\caption{Enclosed angular momentum within $r=0.6\rs0$ for model R5\_RAN (compare with Figure~\ref{f:PE}). The appearance of spiral
modes induces a steady growth in the magnitude of the $L_x$ and $L_z$ components. \emph{Note that the model is evolved to later
times}.
}
\label{f:angmom_encl_rand}
\end{figure}

The effect of changing the initial perturbation is shown in Figure~\ref{f:angmom_encl_rand}, which shows the angular momentum
enclosed within $r = 0.6\rs0$ for model R5\_RAN. 
The angular momentum amplification is directly related to the growth of spiral modes (compare with Figure~\ref{f:PE}).
By the time integration stops, $Y_1^1$ is still growing, so $L_y$ may increase even more in magnitude. The fact that the
spin-up along the z-axis is positive (in contrast to models R5\_L11\_HR and R5\_L11P2c) is a consequence of the
random initial relative phase between the corresponding sloshing modes. The spherical symmetry of the problem
implies that changing the sign of $\Phi$ in equation~(\ref{eq:trigger_eqn}) leads to angular momentum redistribution
with the opposite sign.

The order-of-magnitude of the maximum angular momentum redistribution, $\dot{M}\rs0^2$, can be understood if at saturation, the azimuthal velocity $v_\phi$
is close to the upstream velocity $v_1$. From linear theory (\ref{s:background}), this follows for shock displacements of order unity, which is
indeed the case for $\ell=1$ modes. Not so straightforward to explain is the size of the fluctuations in the fully saturated
state, which involves interference between modes with different amplitudes and phases.
The basic behavior of the spin-up during the exponential growth phase agrees with what was 
seen by \citet{blondin07b} (their Figure~7), although
they did not quantify their results in terms of fundamental physical quantities.
Our results agree to within a factor of a few with those of \citet{blondin07a}. They report typical spin-ups of
$2.5\times 10^{47}$ at 250~ms (72$\tff0$ for $M=1.2M_\sun$ and $\rs0 = 230/1.5 = 153$~km, where 1.5 is our
approximate average shock radius in units of $\rs0$ and $230$~km their quoted shock position). At similar times, we find
$0.6\dot{M}\rs0^2 \simeq 10^{47}(\dot{M}/0.36\,M_\sun~{\rm s}^{-1})(\rs0/153~{\rm km})^2$~g~cm$^2$~s$^{-1}$. The differences could be caused by
several factors: (i) their absorbing boundary condition with no cooling, which causes the shock to expand with time (e.g., \citealt{BM03}), 
and (ii) resolution, which we have already shown to affect the long term behavior of the system.

\begin{figure}
\includegraphics*[width=\columnwidth]{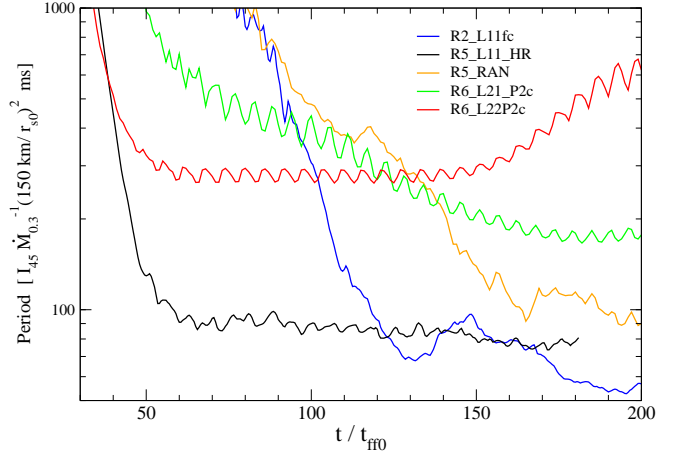}
\caption{Minimum neutron star spin periods, obtained by assuming that all of the angular momentum enclosed within the radius
at which $l_z(r)$ changes sign is accreted onto a star of moment of inertia $I_{\rm NS} = 10^{45}I_{45}$, for different 
models and as a function of time. All components
of the angular momentum are included.
}
\label{f:periods_combined}
\end{figure}

Assuming that all of the angular momentum enclosed within the radius where $l_z(r)$ changes sign is accreted 
onto the neutron star, and that the latter has a moment of inertia 
$I_{\rm NS} = I_{45}\times 10^{45}$~g~cm$^2$, one can write the minimum period due to spiral modes as
\begin{equation}
P \simeq 80\, I_{45} \dot{M}^{-1}_{0.3}\left(\frac{150~{\rm km}}{\rs0} \right)^2 \left(\frac{0.6}{f_{\rm amp}} \right)~{\rm ms},
\end{equation}
where $f_{\rm amp}$ is the fraction of $\dot{M}\rs0^2$ achieved during the phase of exponential growth. If no exponentially
growing spiral mode takes place, one can still expect to achieve $f_{\rm amp}\lesssim 0.1$ from stochastic fluctuations,
in which case the spin period $P\gtrsim 500$~ms. Figure~\ref{f:periods_combined} shows the evolution of the spin period for a 
few representative modes, taking into account the three components of the angular momentum, 
$P = 2\pi I_{\rm NS} / \sqrt{L_x^2+L_y^2+L_z^2}$,
with the $L_i$ measured at the radius at which $l_z(r)$ changes sign.

This simplistic picture is bound to change when more realistic physics is included. First, because it is not clear yet that spiral
modes can easily develop in a convective gain region (e.g., \citealt{iwakami08}), but also because both the neutrinosphere
radius, the mass accretion rate, and the shock radius change as a function of time. The onset of explosion
further complicates the picture, as the mass accretion is cut off and the postshock flow expands. Still, anisotropic cold
and fast downflows piercing the convective region are expected to impart stochastic torques to the forming neutron star \citep{thompson00},
an effect that we cannot capture with our current setup.

\section{Summary and Discussion}

In this paper we have studied the spiral modes of the SASI
in the linear  and nonlinear regime, when no rotation is imposed on the
accretion flow, and when neutrino driven convection is suppressed. 
We have combined results from linear stability analysis and time-dependent 
simulations to understand the structure and evolution
of these non-axisymmetric modes. Our main findings follow:
\newline

\noindent 1. --  Spiral modes are most easily understood as two or more sloshing modes 
out of phase. In the linear regime and in the absence of rotation, their three-dimensional structure can 
be obtained by linear superposition of known axisymmetric eigenfunctions.
The parameters that determine a spiral mode are the number of sloshing
modes involved, their relative phases, and amplitudes.
Two modes of equal amplitude and relative phase equal to
$\pm \pi/2$ is a limiting case of a more general class of non-axisymmetric
mode.
\newline

\noindent 2. -- As long as the initial relative phase is not zero, 
spiral modes survive in the nonlinear phase. For the $\ell=1$, $m=\pm 1$
case, they seem to reach some type of equilibrium relative phase
once all modes are excited in the fully saturated stage. This equilibrium
phase does not seem to depend on the initial phase shift. Hence, the range of
perturbations needed to excite spiral-like behavior leading to 
protoneutron star spin-up is broader than that needed to achieve spiral
modes with a relative phase of $\pi/2$. The absolute sign of the
angular momentum component in the inner and outer region depends on the
sign of the initial relative phase.
\newline

\noindent 3. -- The angular momentum redistribution in the linear and weakly
nonlinear phase of an isolated spiral mode is caused by the spatial dependence
of the angular momentum density, which consists of at least two counterotating 
regions. This division occurs because the eigenmodes have at least one radial
node in the transverse velocity profile (see, e.g., Figure~\ref{f:angmomdens_eigenfunctions}b-c). 
The angular momentum density is non-oscillatory, and increases in magnitude
at nearly twice the growth rate of the spiral mode. For $\ell=1$, we
have found that the dominant term involves a second order
perturbation to the azimuthal velocity, which couples to the background density. 
We see no apparent relation between the formation of the triple point 
at the shock and the angular momentum redistribution, other than a possible 
role in the saturation of the primary mode and hence a cutoff in the growth
of the spin-up.
\newline

\noindent 4. -- The bulk of the angular momentum redistribution is achieved
at the end of the phase of exponential growth. After $200$ dynamical
times, our models achieve a maximum spin-up of at least $0.6\dot{M}\rs0^2$
along the axis of the primary spiral mode, independent
of the resolution employed, but restricted to $\ell=1$. Modes with
$\ell=2$ also lead to spin-up along the primary axis, but with a smaller magnitude.
\newline

\noindent 5. --  In the fully saturated stage, where all modes are excited
due to nonlinear coupling, all components of the angular momentum
fluctuate with a characteristic magnitude $\lesssim \dot{M}\rs0^2/10$. Resolution
seems to be relevant for capturing the long-term evolution of the spin-up, which
can display secular increases in magnitude at low resolution.
\newline

\citet{blondin07a} found that the development of spiral modes at late times is a robust feature
of the flow, as many different types of perturbation resulted in a similar outcome
when the accretion flow is non-rotating. Our results tend to confirm this picture.
This was not the case in the simulations of \citet{iwakami09a}, which however include
neutrino heating and thus convection, fundamentally altering the flow dynamics relative
to the simpler case with no heating. 
The fact that most progenitor models are likely to show some degree of rotation will always
make excitation of a prograde spiral mode more likely, as its growth rate is
larger \citep{yamasaki08}. Indeed, the development of SASI-like spiral modes has
been observed in time dependent studies of accretion disks around a Kerr black hole \citep{nagakura09}

Are overdense shells with large angular extent realistic? Two- and three-dimensional 
compressible simulations of carbon and oxygen shell burning in a $23M_\sun$ star
find that density fluctuations of order $10\%$ are obtained due to internal
waves excited in the stably stratified layers that lie in between convective shells
\citep{meakin06,meakin07a,meakin07b}. These large amplitude fluctuations are
due to the strong stratification, and are correlated
over large angular scales \citep{meakin07a}. The oxygen-carbon interface
lies too far out in radius to reach the stalled shock during the crucial 
few hundred milliseconds after bounce. But if a similar
phenomenon were to occur above the Si shell, it would have
a definite impact on the evolution of the stalled shock.
Given that typical SASI periods are $\sim 30$~ms, all that is needed is a 
region of radial extent 
$\lesssim v_{\rm ff}/\omega_{\rm osc}\sim 200M_{1.3}^{1/2}r_{\rm 150}^{-1/2}$~km
by the time it reaches the shock.

A different question is whether this coherent superposition of 
linear modes can take place in an environment where turbulent
neutrino-driven convection already operates. The convective growth
time in the gain region is a few ms (e.g., \citealt{fryer07}) compared
to the several tens of ms required for the SASI to grow. 
A critical parameter determining the interplay between these two instabilities is the integral of the buoyancy
frequency multiplied by the advection time over the gain region \citep{foglizzo06}.
If this dimensionless parameter $\chi$ is less than 3, then
infinitesimal perturbations do not have time to grow before they are advected
out of the heating region, and a large amplitude perturbation is required
to trigger convection. Using axisymmetric simulations with approximate neutrino
transport and a realistic equation of state, \citet{scheck08} found that 
for weak neutrino heating, the SASI overturns can actually trigger convection.
On the other hand, using parametric simulations, \citet{FT09b}
found that when $\chi > 3$, convection grows rapidly and the vorticity distribution
reaches its asymptotic value before the SASI achieves significant amplitudes.
Two-dimensional convection is volume filling, as the vorticity
accumulates on the largest spatial scales, hence large scale convective modes
could be exciting dipolar SASI modes. The interplay between these two instabilities
is not well understood at present, hence discussion of spiral modes in this
context will have to wait for further work.

A few three-dimensional core-collapse simulations have been performed so far, some
of these with highly dissipative numerical methods \citep{fryer07,iwakami08}.
These groups find that convection starts on smaller angular scales, with convective cells having
roughly the size of the gain region \citep{fryer07}, or being mediated by
high-entropy bubbles with a range of sizes \citep{iwakami08}. In exploding simulations, as the shock 
expands, convective cells increase size and a global $\ell=1$ mode emerges \citep{fryer07,iwakami08}.
Persistent spiral modes do not seem to appear spontaneously, they need to be explicitly
triggered \citep{iwakami08}.

Very recently, \citet{nordhaus10a} have reported three-dimensional,
parametric core-collapse simulations using Riemann solvers and covering the whole sphere. Their results are
in line with previous studies in that they do not witness the development of large scale shock oscillations 
in the stalled phase, or spiral modes with noticeable amplitudes. 
Similarly, \citet{wongwathanarat2010}
have presented full-sphere parametric explosions with a Riemann hydrodynamic solver, including
the contraction of the protoneutron star and grey neutrino transport. Their conclusions are
similar to that of \citet{nordhaus10a} in that no coherent spiral modes are observed, with
angular momenta saturating at a few times $10^{46}$~g~cm$^2$~s$^{-1}$.
\citet{fryer07} observe specific angular momenta $\lesssim 10^{13}$~cm$^2$~s$^{-1}$ imparted 
stochastically to the protoneutron star by anisotropic accretion. 
These values, and those of \citet{wongwathanarat2010}, are in agreement with the angular momentum 
fluctuations we observe in our fully saturated SASI, which on average have a magnitude 
$\sim 5\times 10^{12} (f_{\rm amp}/0.1)\dot{M}_{0.3}(\rs0/150~{\rm km})^2 M^{-1}_{1.3}$~cm$^2$~s$^{-1}$.

\acknowledgements

I am grateful to Aristotle Socrates for discussions on angular momentum from linear modes, and to Thomas
Janka for pointing out the cutout at the polar boundary as a fix to the numerical instability. I also thank Adam Burrows 
and Chris Thompson for constructive comments on the early manuscript. For useful discussions, I thank Tobias Heinemann, 
Shane Davis, Douglas Rudd, Jason Nordhaus, Manou Rantsiou, Brian Metzger, and Tim Brandt. 
The anonymous referee made constructive comments that have resulted in a significantly improved manuscript.
The author is supported by NASA through Einstein Postdoctoral Fellowship
grant number PF-00062, awarded by the Chandra X-ray Center, which is operated
by the Smithsonian Astrophysical Observatory for NASA under contract NAS8-03060.
This research was supported in part by the National Science Foundation through 
TeraGrid resources \citep{catlett07}, provided by NCSA. Computations were performed at the NCSA Abe
and IAS Aurora clusters. I also thank CITA for access to their computational resources.

\appendix

\section{Calculation of Linear Eigenmodes}
\label{sec:v_trans}

\subsection{Background Flow and Perturbation Equations}
\label{s:background}

The steady accretion flow below the shock is obtained by solving the time-independent Euler equations,
using an ideal gas equation of state of adiabatic index $\gamma$ and the gravity of a point mass $M$,
\begin{eqnarray}
\frac{\partial}{\partial r} (r^2 \rho v_r) & = & 0 \\
v_r \frac{\partial v_r}{\partial r} + \frac{1}{\rho}\frac{\partial p}{\partial r} + \frac{GM}{r^2} & = & 0\\
\frac{\partial p}{\partial r} - c_s^2 \frac{\partial \rho}{\partial r} & = & (\gamma-1)\frac{\mathscr{L}}{v_r}.
\end{eqnarray}
The boundary conditions at $r = \rs0$ are given by the Rankine-Hugoniot jump conditions 
(e.g., \citealt{landau}). The upstream flow is adiabatic and has Mach number $\mathcal{M}_1$ at $r=\rs0$. For a given
ratio $r_*/\rs0$, the normalization of the cooling function is obtained by imposing $v_r(r_*) = 0$.

\citet{F07} introduce Eulerian perturbations of the form shown in equation~(\ref{eq:perturbation_form}). To
obtain a compact formulation of the system, the following perturbation variables are employed
\begin{eqnarray}
\label{eq:h_def}
h & = & \frac{\delta v_r}{v_r} + \frac{\delta \rho}{\rho}\\
f & = & v_r \delta v_r + \frac{\delta c^2}{\gamma-1}\\
\delta S & = & \frac{1}{\gamma-1}\left(\frac{\delta p}{p} - \gamma\frac{\delta \rho}{\rho} \right)\\
\label{eq:K_def}
\delta K & = & r^2 \mathbf{v}\cdot (\nabla \times \delta\mathbf{w}) + \ell(\ell+1)\frac{p}{\rho}\delta S,
\end{eqnarray}
corresponding to the perturbed mass flux, energy flux, entropy, and an entropy-vortex combination, respectively.
The differential system for the radial profile of linear perturbations is obtained by perturbing
the time-dependent fluid equations to first order, and rewriting the resulting equations in terms
of (\ref{eq:h_def})-(\ref{eq:K_def}), obtaining \citep{F07}
\begin{eqnarray}
\label{eq:h_evol}
\frac{\partial h}{\partial r} & = & \frac{i\delta K}{\omega r^2 v_r} + \frac{i\omega}{v_r(1-\mathcal{M}^2)}
                                    \left[\frac{\mu^2}{c^2}f -\mathcal{M}^2h -\delta S \right]\\
\frac{\partial f}{\partial r} & = & \delta \left( \frac{\mathscr{L}}{\rho v_r}\right) 
     + \frac{i\omega v_r}{(1-\mathcal{M}^2)}\left[h - \frac{f}{c^2} 
             +\left( \gamma-1 + \frac{1}{\mathcal{M}^2}\right)\frac{\delta S}{\gamma}\right]\\
\frac{\partial \delta S}{\partial r} & = & \frac{i\omega}{v_r} \delta S 
                    + \delta \left( \frac{\mathscr{L}}{p v_r}\right)\\
\label{eq:K_evol}
\frac{\partial \delta K}{\partial r} & = & \frac{i\omega}{v_r}\delta K 
                    + \ell(\ell+1)\delta \left( \frac{\mathscr{L}}{\rho v_r}\right),
\end{eqnarray}
where $\mu^2 = 1 - (1-\mathcal{M}^2)[\ell(\ell+1)c^2/r]/\omega^2$ and $\mathcal{M}$ is the Mach number.
The shock boundary conditions for equations~(\ref{eq:h_evol})-(\ref{eq:K_evol}) are obtained from the 
perturbed shock jump conditions evaluated at the unperturbed shock position $\rs0$ \citep{F07},
\begin{eqnarray}
h_2 & = & \left(\frac{1}{v_2} - \frac{1}{v_1} \right)\Delta v\\
f_2 & = & \left(\frac{\mathscr{L}_1}{\rho_1 v_1} - \frac{\mathscr{L}_2}{\rho_2 v_2}\right)\Delta \xi
           +(v_2 - v_1)\Delta v\\
\delta S_2 & = & \frac{\rho}{p}\left[\frac{\mathscr{L}_1}{\rho_1 v_1} - \frac{\mathscr{L}_2}{\rho_2 v_2} 
                 - \left(\frac{GM}{\rs0^2} - 2\frac{v_1 v_2}{\rs0} \right)\left(1 - \frac{v_2}{v_1}\right)
                \right]\Delta\xi \nonumber\\
           &    & - \frac{v_1}{p/\rho}\left(1 - \frac{v_2}{v_1} \right)^2 \Delta v\\
\delta K_2 & = & \ell(\ell+1)\left[ (v_1 - v_2)\Delta v + f_2\right],
\end{eqnarray}
where $\Delta \xi$ is the shock displacement, $\Delta v = -i\omega \Delta \xi$ the shock velocity, and
the subscripts $1$ and $2$ refer to quantities above and below the shock, respectively.
The reflecting boundary condition at $r = r_*$ is
\begin{equation}
\delta v_r = \frac{v_r}{(1 - \mathcal{M}^2)}\left[h + \delta S - \frac{f}{c^2} \right] = 0,
\end{equation}
resulting in a complex eigenvalue $\omega$.

\subsection{Transverse Velocity Perturbation} 

The transverse components of the perturbed Euler equation are
\begin{eqnarray}
\label{eq:v_theta_Euler}
-i\omega\,\delta v_\theta + \frac{v_r}{r}\frac{\partial\,\delta v_r}{\partial \theta} +
                          v_r\, \delta w_\phi + \frac{1}{\rho\, r}\frac{\partial \delta p}{\partial\theta} & = & 0\\
\label{eq:v_phi_Euler}
-i\omega\,\delta v_\phi + \frac{v_r}{r\,\sin\theta}\frac{\partial\,\delta v_r}{\partial \phi} -
                    v_r\, \delta w_\theta + \frac{1}{\rho\, r\, \sin\theta}\frac{\partial \delta p}{\partial\phi} & = & 0,
\end{eqnarray}
from which one can readily solve for $\delta v_\theta$ and $\delta v_\phi$ if the transverse components
of the vorticity $\mathbf{w}$ are known. Note that, since $\delta v_r$ and $\delta p$ are proportional
to $Y_\ell^m$ (\S\ref{sec:linear_stability}), one has 
$\{\delta v_\theta,\, \delta w_\phi\}\propto \partial Y_\ell^m / \partial \theta$, and 
$\{\delta v_\phi,\, \delta w_\theta\}\propto \partial Y_\ell^m/(\sin\theta\,\partial \phi)$.
Hence the only additional quantity required is the radial profile of the vorticity.

Taking the curl of the Euler equation yields
\begin{equation}
\frac{\partial \delta \mathbf{w}}{\partial t} - \nabla \times \left(\mathbf{v}\times \delta \mathbf{w}\right) -
\frac{1}{\rho^2}\left(\nabla\rho \times \nabla p\right) = 0.
\end{equation}
The inertial term can be expanded as
\begin{equation}
\nabla\times (\mathbf{v}\times\delta\mathbf{w}) = v_r\left(\nabla_\Omega \cdot \delta \mathbf{w}_\Omega\right)\hat r
                                               -\frac{1}{r}\frac{\partial}{\partial r}
                                               \left(r\,v_r\,\delta \mathbf{w}_\Omega \right), 
\end{equation}
where $\nabla_\Omega \cdot \delta\mathbf{w}_\Omega$ is the angular part of the divergence of the transverse vorticity,
$\delta \mathbf{w}_\Omega = \delta w_\theta\, \hat\theta + \delta w_\phi\,\hat\phi$.
Denoting by $\mathbf{B}$ the baroclinic term, one obtains
\begin{eqnarray}
B_r      & = & 0\\
\label{eq:baro_t}
B_\theta & = & -\frac{1}{\rho_0^2}\left[ \frac{1}{r\sin\theta} \frac{\totd p_0}{\totd r}
                                                         \frac{\partial\,\delta\rho}{\partial \phi}
                   -\frac{1}{r\sin\theta} \frac{\totd \rho_0}{\totd r} \frac{\partial\,\delta p}{\partial\phi}\right]\\
\label{eq:baro_f}
B_\phi   & = & -\frac{1}{\rho_0^2}\left[ \frac{1}{r}\frac{\totd \rho_0}{\totd r}
                                       \frac{\partial\, \delta p}{\partial \theta}\phantom{\frac{1}{\sin\theta}}
       -\frac{1}{r} \frac{\totd p_0}{\totd r} \frac{\partial\,\delta\rho}{\partial\theta}\right].
\end{eqnarray}
The equation governing the radial profile of the transverse vorticity is then
\begin{equation}
\label{eq:vorticity_profile_ODE}
\frac{\partial\,\delta\mathbf{w}_\Omega}{\partial r} =  \left[\frac{i\omega}{v_r} + \frac{1}{r} 
                             + \frac{1}{\rho_0}\frac{\totd \rho_0}{\totd r}\right]\delta \mathbf{w}_\Omega
                             -\frac{1}{v_r}\mathbf{B} 
\end{equation}
Once the radial profile of $\delta\mathbf{w}_\Omega$ is known, solving for $\delta w_r$ is straightforward,
as no radial derivatives are involved.

The boundary condition at the shock for equation~(\ref{eq:vorticity_profile_ODE}) is obtained from 
equations~(\ref{eq:v_theta_Euler}) and (\ref{eq:v_phi_Euler}), once the transverse velocity
below the shock is known. Imposing $\hat t_i \cdot (\mathbf{v}_2 
+ \delta\mathbf{v}_2 - \mathbf{v}_1) = 0$ at the shock, where $\hat t_i$ are the tangent vectors
\begin{eqnarray}
\hat t_\theta & = & \frac{1}{r}\frac{\partial \Delta \xi}{\partial\theta}\hat r + \hat\theta\\
\hat t_\phi   & = & \frac{1}{r\sin\theta}\frac{\partial\Delta\xi}{\partial\phi}\hat r + \hat\phi,
\end{eqnarray} 
one obtains (e.g., \citealt{F07})
\begin{eqnarray}
\delta v_{\theta,2} & = & \frac{v_{r,1} - v_{r,2}}{\rs0}\frac{\partial \Delta\xi}{\partial \theta} \\
\delta v_{\phi,2}   & = & \frac{v_{r,1} - v_{r,2}}{\rs0\sin\theta}\frac{\partial \Delta \xi}{\partial\phi},
\end{eqnarray}
where the subscripts $1$ and $2$ denote values upstream and downstream of the shock, respectively. 
We then obtain
\begin{eqnarray}
\label{eq:bnd_wr}
\delta \tilde{w}_{r,2}       & = & 0\\ 
\label{eq:bnd_wt}
\delta \tilde{w}_{\theta,2} & = & -\frac{i\omega}{v_r}\delta\tilde{v}_{\phi,2}+\frac{1}{\rs0}\delta\tilde{v}_{r,2}
                                  + \frac{1}{\rho \rs0}\delta\tilde{p}_2\\
\label{eq:bnd_wf}
\delta \tilde{w}_{\phi,2}   & = & \frac{i\omega}{v_r}\delta\tilde{v}_{\theta,2}-\frac{1}{\rs0}\delta\tilde{v}_{r,2}
                                  - \frac{1}{\rho \rs0}\delta\tilde{p}_2,
\end{eqnarray}
where tildes denote the radial amplitude (eq.~[\ref{eq:perturbation_form}]). Given that $\delta\tilde{v}_{\theta,2} = \delta\tilde{v}_{\phi,2}$, it follows that $\delta\tilde w_{\phi,2} = -\delta\tilde w_{\theta,2}$. With
this boundary condition, equations~(\ref{eq:baro_t}), (\ref{eq:baro_f}), and (\ref{eq:vorticity_profile_ODE})
imply that $\delta \tilde w_\theta(r) = -\delta \tilde w_\phi(r)$ for all $r$, and hence 
$\delta \tilde{v}_\theta(r) = \delta\tilde{v}_\phi(r)$ as well. For convenience, we adopt the notation
\begin{equation}
\delta \tilde{v}_\Omega(r) \equiv \delta \tilde{v}_\theta(r) = \delta\tilde{v}_\phi(r).
\end{equation}
Given the radial and angular form of the transverse velocity perturbations, one obtains
$\delta w_r(r) = (\nabla \times \mathbf{v})_r = 0$ everywhere.

Knowing that the transverse components of the vorticity have equal and opposite radial amplitudes,
one can show from the definition of $\delta K$ (\S\ref{sec:linear_stability}) that
\begin{equation}
\label{eq:K_redef}
\delta \tilde{w}_\theta = \frac{1}{r v_r}\left[\frac{\delta\tilde{K}}{\ell(\ell+1)} - 
                          \frac{p}{\rho}\delta \tilde{S} \right],
\end{equation}
that is, the amplitude of the vorticity can be directly obtained from the basic perturbation
variables, without the need for solving equation~(\ref{eq:vorticity_profile_ODE}).
We have nevertheless integrated this equation as a self-consistency check.

\section{Zeus-MP Implementation}
\label{sec:zeus}

\subsection{Tensor Artificial Viscosity}

Astrophysical hydrodynamic codes that employ finite difference algorithms to solve the Euler
equations rely on artificial viscosity to broaden shocks over a few cells. This allows
differencing of the flow variables without discontinuities, while obtaining the correct solutions
to the jump conditions away from shocks. The standard practice is to use
the prescription of \citet{vonneumann50}, which only takes into account the spatial 
derivative of the velocity normal to the shock as an approximation to the divergence of the
velocity field for determining compression. When curvilinear coordinates are used,
however, this approach results in short wavelength oscillations behind the shock,
and spurious heating of homologously contracting flows (e.g., \citealt{tscharnuter79}).
To fix this, the artificial viscosity can be formulated as a coordinate-invariant tensor,
which can correctly identify zones where the flow is compressed by becoming active only
when $\nabla\cdot \mathbf{v} <0$ \citep{tscharnuter79}. Additional constraints on the
functional form come from requiring that the artificial viscosity does not act on homologously
contracting or shear flows \citep{tscharnuter79,stone92,hayes06}. 

In the context of core-collapse calculations, a tensor form of the artificial viscosity
in Zeus-MP has previously been employed by \citet{iwakami08}, \citet{iwakami09a}, and
\citet{iwakami09b}. They found that unless the tensor formulation is used, the carbuncle
instability \citep{quirk94} appears at the poles of the shock \citep{iwakami08b}.

We have independently implemented a tensor artificial viscosity in Zeus-MP along the lines of 
\citet{stone92} and \citet{iwakami08}, with some slight modifications. First, we 
use the volume difference instead of metric coefficient times coordinate difference \citep{stone92}.
This removes singularities at the polar axis whenever the factor $\sin\theta\totd \theta$ appears
in the denominator. Second, we have made use of the traceless nature of the artificial viscosity
tensor to rewrite the $\theta$-component of its divergences as
\begin{eqnarray}
\left(\nabla \cdot \mathbf{Q}\right)_{\,(2)} 
                           & =  & \frac{1}{r\,\sin\theta}\frac{\partial}{\partial \theta}\left(\sin\theta\,
                               Q_{22}\right) - \frac{Q_{33}}{r\sin\theta}
                               \frac{\partial \sin\theta}{\partial \theta}\\
                           & = & \frac{1}{r\,\sin^2\theta}\frac{\partial}{\partial \theta}
                              \left(\sin^2{\theta}\,Q_{22}\right) + \frac{Q_{11}}{r\sin\theta}
                              \frac{\partial\sin\theta}{\partial \theta},
\end{eqnarray}   
where the second expression is that used by \citet{stone92} and \citet{iwakami08}. This allows to
completely eliminate the metric coefficients of the denominator, by implementing the volume
difference described above. The net effect is to reduce the noise in the $\theta$ velocity
in the supersonic regions close to the z-axis, without making any significant difference at or below the shock.

When evolving our setup with this artificial viscosity prescription, the flow remains
spherically symmetric and smooth for an indefinite time as long as there are no perturbations.
If, on the other hand, the \citet{vonneumann50} prescription is used, the carbuncle instability
indeed appears out of numerical noise within $\sim 10$ dynamical times, in agreement with \citet{iwakami08b}.

However, when evolving sloshing SASI modes transverse to the z-axis, a runaway
sawtooth instability of the $\phi$ component of the velocity develops around the axis
whenever the amplitude (and thus the flow velocity) becomes large. This happens because
the radial velocity at both sides of the axis is not the same, which in turn is caused
by a wake generated by the reflecting axis. The reflection of the flow at the axis
also compresses it, resulting in enhanced artificial viscosity dissipation
at the axis, with irregular velocities. A similar phenomenon is observed in modes
parallel to the z-axis, although the cause of this is not clear yet.

The cutout at the polar axis, described in the next subsection, suppresses the growth of this numerical instability
to the point where meaningful results can be obtained. It does not, however, completely eliminate it. We proceed to
the nonlinear phase based on the fact that the angular momentum contributed by regions around the shock and 
close to the polar axis is minor.

\begin{figure*}
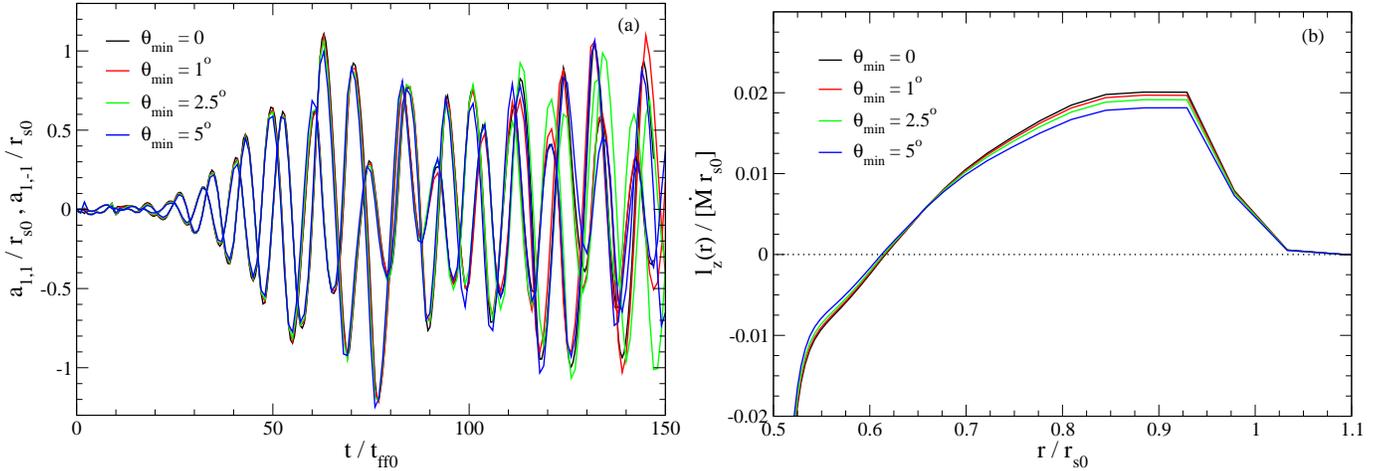

\includegraphics*[width=0.5\textwidth]{f18a.eps}
\includegraphics*[width=0.5\textwidth]{f18b.eps}
\caption{\emph{Left:} spherical harmonic coefficients for modes $Y_1^1$ and $Y_1^{-1}$ as a function
of the half-opening angle $\theta_{\rm min}$ of the cutout around the polar axis. The cases $\theta_{\rm min}=0$
and $\theta_{\rm min} = 5$~deg correspond to models R5\_L11P2 and R5\_L11P2c, respectively. \emph{Right:} z-component
of the angular momentum density integrated over a spherical surface (eq.~[\ref{eq:angular_mom_density}]) at $t=24\tff0$
as a function of $\theta_{\rm min}$, for the same runs as the left panel.}
\label{f:alm_cutout}
\end{figure*}

\begin{figure*}
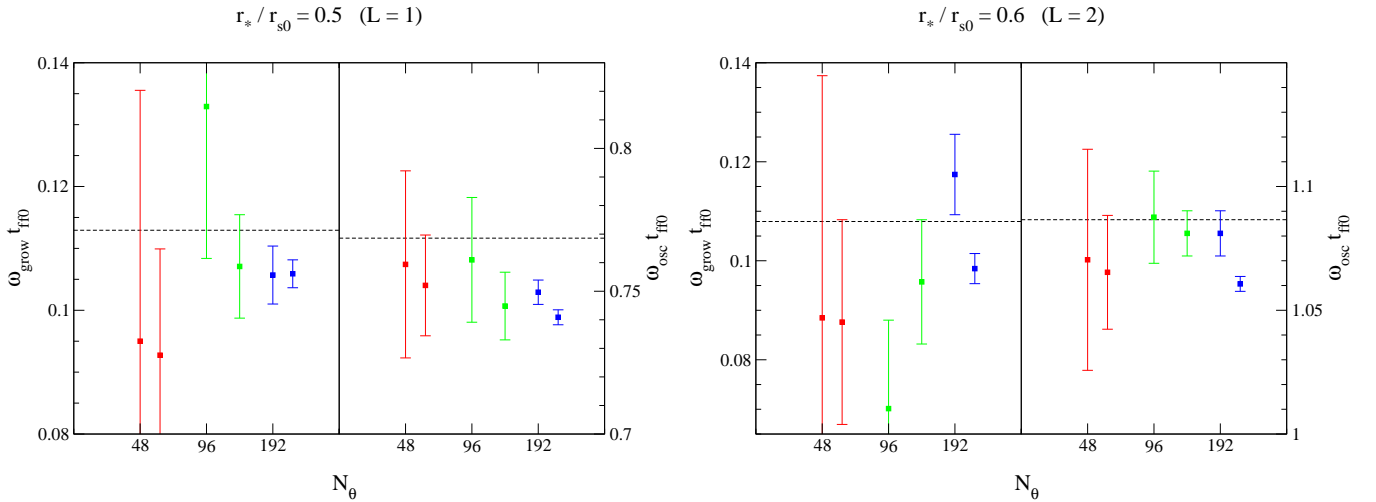

\includegraphics*[width=0.5\textwidth]{f19a.eps}
\includegraphics*[width=0.5\textwidth]{f19b.eps}
\caption{Comparison of linear growth rates (left panels) and oscillation frequencies (right panels) 
from numerical simulations (squares) with values from linear stability analysis (dashed lines). The
left plot shows results for $\ell=1$ ($r_*/\rs0 = 0.5$) and the right $\ell=2$ ($r_*/\rs0 = 0.6$).
Different angular resolutions are color coded, with red, green, and blue corresponding to $N_\theta=48$,
$96$ and $192$ cells over the whole range of polar angles, respectively. For each angular resolution, two points
are shown, one with the baseline radial resolution (such that $\rs0\Delta \theta = \Delta r$ at the shock)
and another (to the right) with radial resolution doubled. At the resolution employed in this study,
growth rates are reproduced within $20\%$, and oscillation frequencies within $10\%$ of the
linear stability value.}
\label{f:figure_restest}
\end{figure*}

\subsection{Cutout at the Polar Axis}

To investigate the reliability of results obtained with the axis cutout prescription described in \S\ref{sec:time_dep_setup},
we performed two runs with half-opening angles $1$ and $2.5$ degrees, which fill the gap between models R5\_L11P2 and R5\_L11P2c.
Figure~\ref{f:alm_cutout}a shows the spherical harmonic coefficients for modes $Y_1^1$ and $Y_1^{-1}$ as a function
of time, for different values of the half-opening angle $\theta_{\rm min}$ of the cone that is removed from the grid around the
polar axis. Aside from a $\sim 10\%$ reduction in the maximum amplitude, all curves follow nearly identical trajectories 
until $t\simeq 80\tff0$, where they start to diverge more noticeably. Panel (b) shows the angular momentum density
integrated over a spherical surface at time $t=24\tff0$ (eq.~[\ref{eq:angular_mom_density}], a low-resolution counterpart
to Figure~\ref{f:angmom_components}). Again, as the opening angle increases, results decrease by about $10\%$, consistent
with the decrease in amplitude of shock oscillations. Other than that, the qualitative behavior is identical. We conclude that
our resolution-independent results are reliable, with a quantitative uncertainty of at least $10\%$.

\subsection{Linear Growth Rates}

The eigenfrequencies obtained with the method described in Appendix~\ref{sec:v_trans} have been
verified to high precision with time-dependent axisymmetric hydrodynamic simulations using
the code FLASH2.5 \citep{FT09a}. A basic test for the newly implemented setup is thus the reproduction
of the linear SASI growth rates from linear stability analysis. 
This is a resolution-dependent requirement, hence convergence tests are needed. 
Figure~\ref{f:figure_restest} shows growth rates and oscillation frequencies as a function
of angular and radial resolution, for axisymmetric simulations of $\ell=1$ and $\ell=2$ modes.
The measurement method is the same as that described in \citet{FT09a}. With the baseline resolution 
employed in this study ($N_\theta = 48$), growth rates are captured within 
$\sim 20\%$ and oscillation frequencies within $10\%$.

\bibliographystyle{apj}
\bibliography{sasi3d,apj-jour}

\end{document}